\documentclass[amsmath,amssymb,aps,pra,reprint,superscriptaddress]{revtex4-2} 
\usepackage{amsmath,amssymb}
\usepackage[english]{babel}
\usepackage{ bbold }
\usepackage{upgreek}
\usepackage{dsfont}
\usepackage{graphicx}
\usepackage{comment}
\usepackage{braket}
\usepackage{colortbl}
\usepackage{blindtext}
\usepackage[hidelinks]{hyperref}
\usepackage{booktabs}
\usepackage{enumitem}

\newcommand{\blc}[1]{\boldsymbol{#1}} 

\newcommand{\tcb}[1]{{#1}} 

\begin{abstract}
We show that matter-wave diffraction off a single standing laser wave can be used as an accurate measurement scheme for photophysical molecular parameters. These include state-dependent optical polarizabilities and photon-absorption cross sections, the relaxation rates for fluorescence, internal conversion, and intersystem crossing, as well as ionization or cleavage probabilities. 
We discuss how the different photophysical processes manifest as features of the interference pattern,
and we determine the accuracy of molecular parameters estimated from a realistic measurement with finite particle numbers.
The analysis is based on an analytic calculation in Wigner representation, which accounts for the laser-induced coherent and incoherent dynamics, for the finite longitudinal and transverse coherence in the matter-wave beam, the gravitational and Coriolis acceleration, and an imperfect standing laser wave.
\end{abstract}

\begin{document}

\title{Probing molecular photophysics in a matter-wave interferometer}

\author{Lukas Martinetz}
\affiliation{University of Duisburg-Essen, Lotharstra\ss e 1, 47058 Duisburg, Germany}
\author{Benjamin A. Stickler}
\affiliation{Ulm University, Institute for Complex Quantum Systems and Center for Integrated Quantum Science and Technology, Albert-Einstein-Allee 11, 89069 Ulm, Germany}
\author{Ksenija Simonović}
\affiliation{University of Vienna, Boltzmanngasse 5, A-1090 Vienna, Austria}
\author{Richard Ferstl}
\affiliation{University of Vienna, Boltzmanngasse 5, A-1090 Vienna, Austria}
\author{Christian Brand}
\affiliation{German Aerospace Center (DLR), Institute of Quantum Technologies, Wilhelm-Runge-Straße 10, 89081 Ulm, Germany}
\author{Markus Arndt}
\affiliation{University of Vienna, Boltzmanngasse 5, A-1090 Vienna, Austria}
\author{Klaus Hornberger}
\affiliation{University of Duisburg-Essen, Lotharstra\ss e 1, 47058 Duisburg, Germany}

\maketitle

\section{Introduction}
Light can induce various electromagnetic processes in complex molecules. Apart from getting polarized, a molecule can absorb photons that change its electronic state. The absorbed energy may be distributed to vibrational states by a radiationless transition that conserves (internal conversion) or changes (intersystem crossing) the total spin. Alternatively, the molecule may relax to its electronic ground state by emitting fluorescence light. If the photon energy is high enough, the molecule might even ionize or cleave. Which of these processes may occur is quantified  by photophysical parameters such as the polarizability, the absorption cross section, and different quantum yields. 

\begin{figure*}[]
    \centering
    \includegraphics[width=0.99\textwidth]{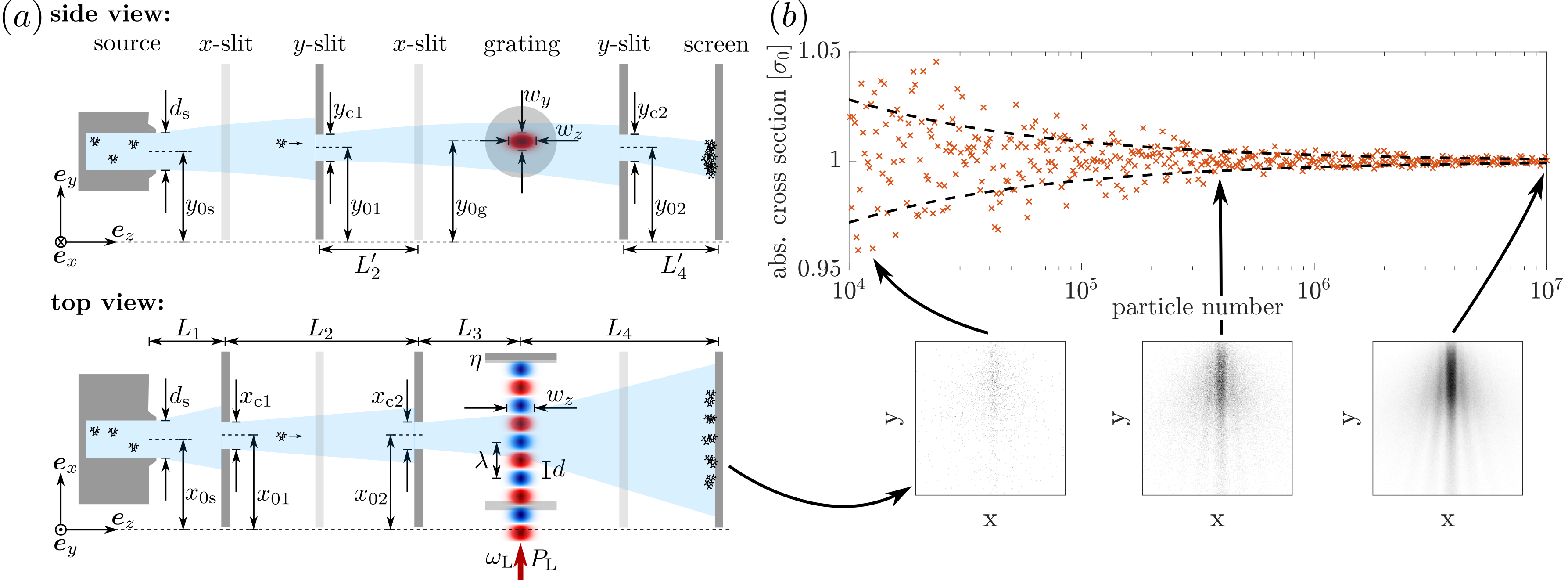}
    \caption{(a) In the interferometric setup considered in this article the molecules are emitted from the source on the left, traverse several collimating elements, interact with a standing laser wave, and are finally detected at a screen on the right. (b) Molecular photophysical parameters such as the absorption cross section determine the shape of the interference pattern (bottom). They can thus be estimated from experimental data by using a formula for the expected 2D pattern. The accuracy of the parameter estimation depends on the number of molecules on the screen (top). This is confirmed by simulating the detection outcomes for different particle numbers and fitting them to the expected fringe pattern (crosses), which verifies our analytical result for the accuracy (dashed lines).  Our calculations and simulations demonstrate that well-characterized state-of-the-art matter-wave interferometers can serve to measure molecular properties in vacuum with high accuracy. Specifically, the above curve indicates that $10^6$ particles would suffice to determine the absorption cross section with an accuracy of about $0.3\%$. The calculation and the iterative fitting procedure are described in Sec.~\ref{sec:iterative_fit_general}. The simulation parameters are motivated by a real-world experiment, as given in App.~\ref{app:simulation_parameters_fig_1}.}
    \label{fig:setup}
\end{figure*}

Photophysical parameters of isolated molecules are often inferred from bulk measurements of the permittivity of solids and liquids, provided the effect of the molecular environment can be corrected for \cite{bonin1997electric}. However,   to validate  theoretical calculations and to determine  gas phase properties, a direct measurement of single molecules flying in vacuum seems more advantageous. 

In vacuum, not only are the internal molecular dynamics accessible, but also the effect of the light on the center of mass of the isolated particle can be studied. Notably, every photon absorption is accompanied by a kick of the molecule momentum, and  a quantum mechanical phase gets imprinted onto the center-of-mass wave function in proportion to the polarizability. 
In modern matter-wave interferometers  \cite{gerlich2007kapitza,haslinger2013,fein2019quantum} such effects are exploited to implement diffraction gratings by means of standing laser waves: the phase imprinted by the periodic optical dipole potential of the standing wave gives rise to a superposition of grating momentum kicks that diffract the molecular matter wave, while every photon absorption further splits the wave packet by half of a grating momentum. In addition, ionization or cleavage at the antinodes of the standing wave can cause the depletion of the  molecular beam, thereby implementing an effective absorption grating. The fringe pattern at the detection screen crucially depends on the rates and quantum yields of the various photophysical processes, rendering it a sensitive probe for the latter.

Interferometric measurement schemes expand established techniques for isolated molecules, ranging from beam deflection experiments \cite{antoine1999direct}, to time-of-flight measurements with particle fountains in static electric fields \cite{amini2003high}, to the observation of spatial patterns due to the classical interaction with a standing laser wave \cite{ballard2000absolute}. Following early applications  of atom interferometry \cite{ekstrom1995measurement}, near-field interference setups with three gratings  were used more recently to measure molecular properties such as  static \cite{berninger2007polarizability,deachapunya2007thermal} or optical polarizabilities \cite{hackermuller2007,gerlich2008matter} and absorption cross sections \cite{eibenberger2014}.

In the present article, we assess how accurately molecular parameters can be measured in a far-field interferometer with a single light grating. Our analysis is motivated in part by recent experiments at the University of Vienna that observed various photophysical signatures of different molecules in such a setup \cite{Simonovic2024}. 
The description unifies models for single-photon ionization \cite{nimmrichter2011}, multi-photon absorption \cite{bateman2014,nimmrichter2014springer,walter2016multiphoton}, with descriptions of internal conversion and intersystem crossing, and fluorescence \cite{walterDiss}. We use 
quantum master equations to describe how the interaction with the laser field entangles the internal molecular state with the center-of-mass motion. The matter-wave interference is formulated by means of the Wigner function, which allows us to account for particle sources with finite extension and velocity spread, all collimation slits, the acceleration due to gravity and the Coriolis force, and a screen not necessarily located in the far field.

\section{Interferometric measurement of photophysical parameters}\label{sec:interferometric_setup}

The basic configuration of the molecule diffraction experiment considered in this work is depicted in Fig.~\ref{fig:setup}(a). 
It consists of a continuous particle source on the left, several collimation and coherence slits, a diffraction grating formed by a standing laser wave, and a spatially resolving two-dimensional detector on the right. The slits in the horizontal direction ($y$-slits)  serve as a velocity selector; they restrict the free-flight parabolas such that slower particles hit the screen closer to the bottom  than faster particles. The slits in the vertical direction ($x$-slits), which are much narrower than the $y$-slits serve to prepare the coherence in the molecular  beam; they restrict the $x$-position of the particles such that several anti-nodes of the standing wave are illuminated coherently. An interference pattern is then formed at the position of the detection screen whose fringe spacing is greater at the bottom than at the top due to the larger de Broglie wave length of the slower particles,  see Fig.~\ref{fig:setup}(b). 
The geometry of the interferometer is explained in detail in Sec.~\ref{sec:interferometer_geometry}.

\begin{figure}
    \centering
    \includegraphics[width=0.49\textwidth]{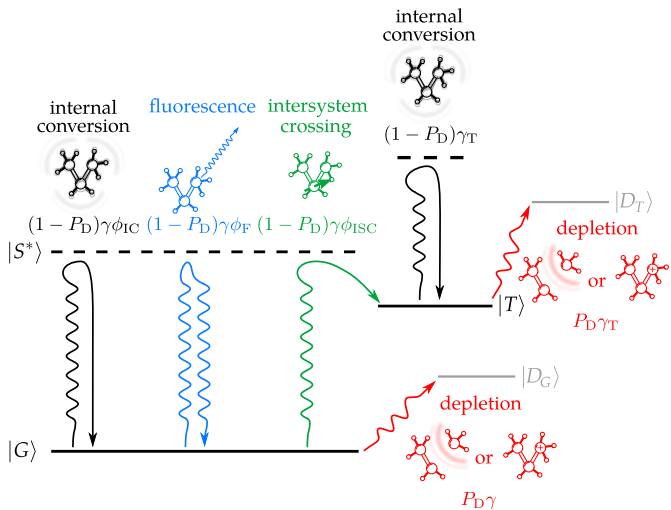}
    \caption{Simplified level scheme used to calculate the Talbot coefficients. It involves a singlet ground state $\ket{G}$, an excited singlet state $\ket{S^*}$ and a triplet state $\ket{T}$. If a photon is absorbed in the ground state, the molecule is excited to the singlet state. Due to the short lifetime of the singlet, the molecule relaxes immediately through either internal conversion (black), fluorescence (blue) or intersystem crossing (green) with the indicated  rates. In addition, the molecule may be ionized or cleaved (red) and thus be removed from the beam, occupying the depletion states $\ket{D_G}$ and $\ket{D_T}$. Note that the lines only indicate the electronic states. After absorbing a photon and relaxing into its electronic ground state, the molecule ends up in a different vibrational state. However, accounting for this change of the vibrational state would yield the same Talbot coefficients as the simplified scheme. \tcb{Note that the polarizabilities and absorption cross sections are here assumed to be independent of the vibrational states.}}
    \label{fig:internal_processes}
\end{figure}

The interaction of the molecules with the laser grating involves a number of photophysical processes, as depicted in Fig.~\ref{fig:internal_processes} and discussed in Sec.~\ref{sec:particle-grating_interaction}. In particular, the absorption of a photon is always accompanied by an excitation (and subsequent deexcitation) of the electronic state. We note that interference is still possible after such a change of internal state, even though only those paths interfere that end up in the same internal state. If the photon absorption leads to ionization or the cleavage of a group of atoms, the molecule is effectively removed from the beam. Since photon absorption is more likely in an anti-node than in a node, this modulates the density in the molecular beam, much like a material grating, and thus contributes to interference at the screen. It follows that all mentioned photophysical processes are encoded in the precise shape of the expected diffraction pattern. In Sec.~\ref{sec:interference_pattern} we provide a formula for the 2D-pattern in the considered setup, see Eq.~\eqref{eq:currentDensity}, as derived in the appendix, and we discuss how the molecular parameters, the laser power, and the slit widths affect the fringe pattern.  

The formula for the diffraction pattern now enables estimating molecular parameters from experimental data. 
The accuracy of such a parameter estimation depends on the number of fit parameters and the number $Z$ of particles detected  on the screen. This is demonstrated in Fig.~\ref{fig:setup}(b) for a single fit parameter, the absorption cross section. We simulate measurement outcomes for different particle numbers and compare the estimated cross sections with both the actual value and with our analytical expression for the expected variance of the estimator. Our calculations and Fig.~\ref{fig:setup}(b) demonstrate that well-characterized state-of-the-art matter-wave interferometers can serve to measure molecular properties with high accuracy. In Sec.~\ref{sec:measuring_molecular_properties} we discuss the accuracy of fitting several molecular parameters simultaneously by two different algorithms. It shows that internal processes of complex molecules can be studied in vacuum. 

\section{Interferometer geometry}\label{sec:interferometer_geometry}

Let us now specify the interferometric setup in more detail, see Fig.~\ref{fig:setup}(a). The particle source emits a continuous stream of molecules of mass $m$. 
They enter the vacuum chamber through a pinhole of 
\tcb{width} $d_{\rm s}$ centered at ($x_{0{\rm s}}$, $y_{0{\rm s}}$). We assume position and momentum to be  uncorrelated at the pinhole, so that the phase space distribution reads
\begin{align}\label{eq:initial_current_density}
    w_{\rm s}(x,y,\blc{p})=& \frac{\mu(\blc{p})}{d_{\rm s}^2}\Theta\!\left(\frac{d_{\rm s}}{2}-|x-x_{0{\rm s}}|\right)\Theta\!\left(\frac{d_{\rm s}}{2}-|y-y_{0{\rm s}}|\right).
\end{align}
Here, $\Theta(\cdot)$ denotes the Heaviside step function. 
The actual form of the pinhole irrelevant because it is masked by the thin $x$-slits in the beam path; only the extension of the source in $y$-direction has an effect on the diffraction pattern. The function $\mu(\blc{p})=\mu_{x}(p_x)\mu_{y}(p_y)\mu_{z}(p_z)$ denotes the momentum distribution of the molecules leaving the source. In the simulations below it will be assumed to be characterized by a source temperature $T$ and a momentum offset $\blc{p}_0=p_{0,z}\blc{e}_z$ to account for a supersonic expansion \cite{morse1996supersonic,scoles1988atomic},$\mu(\blc{p})\propto\exp\!\left[-{(\blc{p}-\blc{p}_{0})^2}/{2mk_{\rm B}T}\right].$

Gravity and the Coriolis force cause an acceleration $a_{{\rm c},x}\blc{e}_x+(g+a_{{\rm c},y})\blc{e}_y$ with $a_{{\rm c},x}\simeq-2\blc{\omega}\cdot\blc{e}_yp_z/m$ and $a_{{\rm c},y}\simeq 2\blc{\omega}\cdot\blc{e}_xp_z/m$. The angular frequency vector $\blc{\omega}$ describes the rotation of Earth. Here we take the dominant momentum component of the particle beam to be $p_z$ and to be unaffected by gravity, the Coriolis force, the slits, and the grating. 

The $x$-slits, with centers  $x_{0i}$ and widths $x_{{\rm c}i}$,  are  placed at distances $L_1$ and $L_1+L_2$ downstream. They serve to collimate the particle beam in $x$-direction to prepare its transverse coherence. Specifically, they ensure that the $p_x$-momentum spread of all molecules passing through both slits is restricted geometrically to a value below the grating momentum. The diffraction at the second  $x$-slit is accounted for in our calculation.

Similarly, two  $y$-slits serve to collimate the beam in the vertical direction, see Fig.~\ref{fig:setup}(a). For infinitely thin slit widths (and in absence of the Coriolis force), they would act as a perfect velocity selector due to gravity, where every height at the screen would correspond to a distinct $p_z$-momentum. In practice, due to the small effect of the Coriolis force and the finite size of the source, a single $y$-slit is already sufficient to correlate the particle velocity and the height at the screen such that the longitudinal coherence suffices to produce  well-separated diffraction peaks. Our calculation accounts for two finite-sized slits to keep the setup general.

The diffraction occurs at a standing laser wave, oriented in $x$-direction, with wavelength $\lambda_{\rm L}$ and grating constant $d=\lambda_{\rm L}/2$. It is placed at a distance $L_3$ behind the second $x$-slit. The laser wave can interact with the molecules through various coherent and incoherent processes, which are fully characterized by the so-called Talbot coefficients $B_n(\xi,y)$ of the grating, as specified in the subsequent section.

\section{Molecule-grating interaction}\label{sec:particle-grating_interaction}
The optical grating is generated by pointing a laser beam with  electric field  ${\rm Re}[\tilde{\blc{E}}(\blc{r})e^{-i\omega_{\rm L} t}]$, towards a retro-reflecting mirror, see Fig.~\ref{fig:setup}. The incident running wave has the complex amplitude $\tilde{\blc{E}}(\blc{r})=E_0\blc{e}_0f(y,z)e^{-ik_{\rm L}x}$, with position vector $\blc{r}=x\blc{e}_x+y\blc{e}_y+z\blc{e}_z$, Gaussian envelope $f(y,z)=\exp\!\left[-(y-y_{0{\rm g}})^2/w_y^2-z^2/w_z^2\right]$, wave vector $k_{\rm L}=\pi/d$, polarization direction $\blc{e}_0\perp \blc{e}_x$, and power $P_{\rm L}=E_0^2\pi\varepsilon_0c w_yw_z/4$. The incident and retro-reflected waves interfere, $\tilde{\blc{E}}(\blc{r})=2E_0\blc{e}_0f(y,z)g(x)$ with 
\begin{equation}\label{eq:definition_gx}
g(x)=\frac{1}{2}\left(\eta\,e^{ik_{\rm L}x}+e^{-ik_{\rm L}x}\right).
\end{equation}
Perfect reflection at the mirror corresponds to $\eta=1$. Realistic ultra-violet mirrors have a finite reflectivity $\eta< 1$.
\subsection{Photophysical processes}

Next, we discuss the various effects occurring when a  complex molecule interacts  with the laser grating.

\subsubsection{Phase grating}
A molecule \tcb{characterized by a} polarizability $\alpha$ \tcb{at the laser frequency is subject } to the (time-averaged) optical dipole potential
\begin{align}
    V(x,y,z)&=-\frac{1}{4}\alpha_{\rm r}\left|\tilde{\blc{E}}(\blc{r})\right|^2\nonumber\\
    &=-\frac{4P_{\rm L}\alpha_{\rm r}}{\pi\varepsilon_0 c w_yw_z}f^2(y,z)|g(x)|^2,\label{eq:potential_definition}
\end{align}
where $\alpha_{\rm r}={\rm Re}(\alpha)$.
In the {eikonal approximation}, the particle trajectories in the laser field are approximately free. After passage through the grating, the molecule has acquired the {eikonal phase} 
\begin{equation}
    \phi(x,y,p_z)=-\frac{1}{\hbar}\int_{-\infty}^\infty dt \,V\!\left(x,y,\frac{p_z t}{m}\right)=\phi_0(y,p_z)|g(x)|^2
\end{equation}
with
\begin{equation}\label{imprinted_phase_def}
    \phi_0(y,p_z)=\sqrt{\frac{8}{\pi}}\frac{P_{\rm L}\alpha_{\rm r}m}{\hbar\varepsilon_0cw_yp_z}\exp\!\left[-2\frac{(y-y_{0{\rm g}})^2}{w_y^2}\right].
\end{equation}
The $x$-dependent phase leaves the spatial distribution of the particles invariant but it modifies their momentum $p_x$. The evolution from the grating to the screen transforms the imprinted momentum state into a spatial diffraction pattern. 

The polarizability may depend on the electronic state of the molecule. We denote the polarizability in the singlet ground state by $\alpha_{\rm r}$ and in the triplet state by $\alpha_{\rm r,T}$, see the level scheme in Fig.~\ref{fig:internal_processes}. The potential $V_{\rm T}$ and the phase $\phi_{0,{\rm T}}$ in the triplet state are defined in analogy to Eqs.~\eqref{eq:potential_definition} and \eqref{imprinted_phase_def}, respectively.

\subsubsection{Photon absorption}
The mean rate $\gamma$ at which photons are absorbed by a molecule
is quantified by the  absorption cross section $\sigma$,
\begin{align}
    \gamma(x,y,z)&=\frac{\sigma}{\hbar \omega_{\rm L}}\frac{\varepsilon_0 c}{2}|\tilde{\blc{E}}(\blc{r})|^2,\nonumber\\
    &=\frac{8P_{\rm L}\sigma}{\pi \hbar \omega_{\rm L}w_yw_z}
    f^2(y,z)|g(x)|^2.\label{eq:absorption_rate2}
\end{align}
In the eikonal approximation, the mean number of photons that have been absorbed after the molecule traversed the grating reads
\begin{equation}\label{eq:number_of_absorbed_photons}
    n(x,y,p_z)=\int_{-\infty}^\infty dt\,\gamma\!\left(x,y,\frac{p_z t}{m}\right)=n_0(y,p_z)|g(x)|^2,
\end{equation}
with
\begin{equation}\label{number_photons_def}
    n_0(y,p_z)=\frac{8}{\sqrt{2\pi}}\frac{P_{\rm L} \sigma m}{\hbar\omega_{\rm L}w_yp_z}\exp\!\left[-2\frac{(y-y_{0{\rm g}})^2}{w_y^2}\right].
\end{equation}
In the quantum mechanical calculation  the function $g(x)$, which describes the position-dependence of the absorption rate, is promoted to an operator $g({\rm x})$, see App.~\ref{app:derivation_Talbot_master_eq}. Applied to the particle state, it effects  a superposition of two opposite momentum kicks $\pm \hbar k_{\rm L}=\pm\hbar \pi/d$, as a result of photon absorption.

\subsubsection{Depletion by photo-cleavage or ionization}
Given that a photon has been absorbed, the molecule will leave the beam with probability $P_{\rm D}\in [0,1]$ due to cleavage or ionization. We assume that $P_{\rm D}$ is independent of the electronic level. However, the rates $\gamma P_{\rm D}$ and $\gamma_{\rm T}P_{\rm D}$ of depletion events can differ for the different levels. In Fig.~\ref{fig:internal_processes} depletion is indicated by the red arrows.

The molecule will stay in the beam with probability $1-P_{\rm D}$. If the molecule is in its ground state, the photon energy can be redistributed by internal conversion (IC), intersystem crossing (ISC), or fluorescence (F) with the respective yields $\phi_i\in[0,1]$,
\begin{equation}
    \phi_{\rm IC}+\phi_{\rm ISC}+\phi_{\rm F}=1.
\end{equation}

\subsubsection{Internal conversion}
The absorbed photon excites the molecule from the ground state to an excited singlet state. \tcb{We assume the  singlet lifetime to be much shorter than the grating interaction time, as is the case for most molecules \cite{walterDiss}. The} 
radiationless transition back to the electronic ground state by exciting vibrational states \tcb{can thus be taken to be immediate}. This process is indicated by the black arrows. 

\subsubsection{Intersystem crossing}
The molecule may also perform a rapid radiationless transition from the ground state via the excited singlet state to the triplet state by flipping a spin. This process is displayed by the green arrow. We neglect the relaxation from the triplet to the ground state through phosphorescence or intersystem crossing because it typically occurs on a much longer timescale than the competing processes. If the molecule absorbs a photon in the triplet state it relaxes through internal conversion.

\subsubsection{Fluorescence}
Another way of relaxing from the excited singlet state to the ground state is by emitting a fluorescence photon, indicated by the blue arrow in Fig.~\ref{fig:internal_processes}. We assume that the distribution of fluorescence photons with momenta $\hbar\blc{k}$ is isotropic. It is characterized by the spectrum $\nu(k)$, where $k=|\blc{k}|$ and $\int d^3k\,\nu(k)/4\pi k^2=1$. Its characteristic function 
\begin{equation}
\varphi(x)=\int \frac{d^3k}{4\pi k^2}e^{-i\blc{k}\cdot \blc{e}_{x}x}\nu(k)
\end{equation}
is used below.

\subsection{Talbot coefficients}
A master equation that accounts for all the mentioned processes, as well as the finite reflectivity of the grating mirror, is provided in App.~\ref{app:derivation_Talbot_coeff}. We solve the dynamics for a molecule initially in the ground state and traversing the Gaussian envelope of the laser grating. The final state after the passage determines the Talbot coefficients
\begin{equation}\label{talbot_coefficients_fluorescence}
B_n(\xi,y)=\frac{1}{d}\int_{-d/2}^{d/2}dx\,e^{-2\pi inx/d}F\!\left(x-\xi\frac{d}{2},x+\xi\frac{d}{2}\right)
\end{equation}
through the transmission function
\begin{align}\label{eq:F_maintext}
F(x,x')=&\frac{\phi_{\rm ISC}N\left(e^{D_{\rm T}+N_{\rm T}}-e^{D+[\phi_{\rm IC}+\phi_{\rm F}\varphi(x-x')]N}\right)}{D_{\rm T}+N_{\rm T}-D-[\phi_{\rm IC}+\phi_{\rm F}\varphi(x-x')]N}\nonumber\\
&+e^{D+[\phi_{\rm IC}+\phi_{\rm F}\varphi(x-x')]N}.
\end{align}
Here, we defined
\begin{subequations}\label{eq:definitions_DN}
\begin{align}
D=&i\left(|g(x)|^2-|g(x')|^2\right)\phi_0(y,p_z)\nonumber\\
&-\frac{1}{2}\left(|g(x)|^2+|g(x')|^2\right)n_0(y,p_z),\\
D_{\rm T}=&i\left(|g(x)|^2-|g(x')|^2\right)\phi_{0{\rm T}}(y,p_z)\nonumber\\
&-\frac{1}{2}\left(|g(x)|^2+|g(x')|^2\right)n_{0T}(y,p_z),\\
N=&(1-P_{\rm D})g^*(x')g(x)n_0(y,p_z),\\
N_{\rm T}=&(1-P_{\rm D})g^*(x')g(x)n_{0{\rm T}}(y,p_z).
\end{align}
\end{subequations}
Note that the Fourier transform in Eq.~\eqref{talbot_coefficients_fluorescence} can be evaluated analytically if $\alpha_{\rm r, T}=\alpha_{\rm r}$, $\sigma_{\rm T}=\sigma$ and $\phi_{\rm F}=0$, see App.~\ref{app:talbot_coefficients}.

\section{The interference pattern}\label{sec:interference_pattern}
The current density at the screen follows from propagating the Wigner function \eqref{eq:initial_current_density} through the entire interferometer, see App.~\ref{app:calculating_pattern},
\begin{align}\label{eq:currentDensity}
    &j(x,y)\propto\int_{-\infty}^\infty dk\,e^{ikx}\sum_{n=-\infty}^\infty\int_0^\infty dp_z\,p_z^2\mu_z(p_z)\nonumber\\
    &\times\chi_0\!\left[(L_3+L_4)\frac{\hbar k}{p_z}-\frac{L_3}{p_z}\frac{2\pi\hbar}{d}n\,,-\hbar k+\frac{2\pi \hbar}{d}n\right]\nonumber\\
    &\times \exp\!\left\{\frac{i}{\hbar}\left[-\frac{1}{2}a_{{\rm c},x}m^2\frac{(L_3+L_4)^2}{p_z^2}\hbar k+\frac{1}{2}a_{{\rm c},x}m^2\frac{L_3^2}{p_z^2}\frac{2\pi\hbar}{d}n\right]\right\}\nonumber\\
    &\times \int_{-\infty}^\infty d\tilde{y}\,h\!\left[y\,,\frac{p_z}{L_4}\left(y-\tilde{y}+\frac{1}{2}(a_{{\rm c},y}+g)\frac{L_4^2m^2}{p_z^2}\right)\,,p_z\right]\nonumber\\
    &\times B_n\!\left(\frac{L_4\hbar k}{dp_z}\,,\tilde{y}\right).
\end{align}
Here, collimation and diffraction of the particle beam at the $x$-slits is accounted for by the function
\begin{align}\label{eq:definitionChi}
    \chi_0(s,q)=&\frac{x_{{\rm c}1} p_z}{L_2}\exp\!\left[\frac{i}{\hbar}\frac{sp_z}{L_2}\left(x_{01}-x_{02}-\frac{1}{2}a_{{\rm c},x}m^2\frac{L_2^2}{p_z^2}\right)\right]\,\nonumber\\
    &\times \exp\!\left(\frac{i}{\hbar}x_{02}q\right){\rm sinc}\!\left[\left(q-\frac{s}{L_2}p_z\right)\frac{x_{{\rm c}2}-|s|}{2\hbar}\right]\nonumber\\
    &\times{\rm sinc}\!\left(\frac{sx_{{\rm c}1}p_z}{2\hbar L_2}\right)(x_{{\rm c}2}-|s|)\,\Theta(x_{{\rm c}2}-|s|),
\end{align}
where ${\rm sinc}(x)=\sin(x)/x$. The finite size of the source and the velocity selection due to the two $y$-slits enter Eq.~\eqref{eq:currentDensity} through
\begin{align}\label{funktionFTwoSlits}
    h(y,p_y&,p_z)=\mu_y\!\left(p_y-\frac{m^2(g+a_{{\rm c},y})L}{p_z}\right)\nonumber\\
    &\times\Theta\!\left(\frac{d_{\rm s}}{2}-\left|y-y_{0{\rm s}}-L\frac{p_y}{p_z}+\frac{(g+a_{{\rm c},y})L^2m^2}{2p_z^2}\right|\right)\nonumber\\
    &\times \Theta\!\left(\frac{y_{{\rm c}2}}{2}-\left|y-y_{02}-L_4'\frac{p_y}{p_z}+\frac{(g+a_{{\rm c},y})L_4'^2m^2}{2p_z^2}\right|\right)\nonumber\\
    &\times\Theta\!\left(\frac{y_{{\rm c}1}}{2}-\left|y-y_{01}-(L_2'+L_3+L_4)\frac{p_y}{p_z}\right.\right.\nonumber\\
    &\left.\left.+\frac{(a_{{\rm c},y}+g)(L_2'+L_3+L_4)^2m^2}{2p_z}\right|\right).
\end{align}
Note that Eq.~\eqref{eq:currentDensity} is independent of the distribution $\mu_x$ because the narrow $x$-collimation restricts $p_x$ to a small interval of equiprobable momenta.
\begin{figure*}[]
    \centering
    \includegraphics[width=0.85\textwidth]{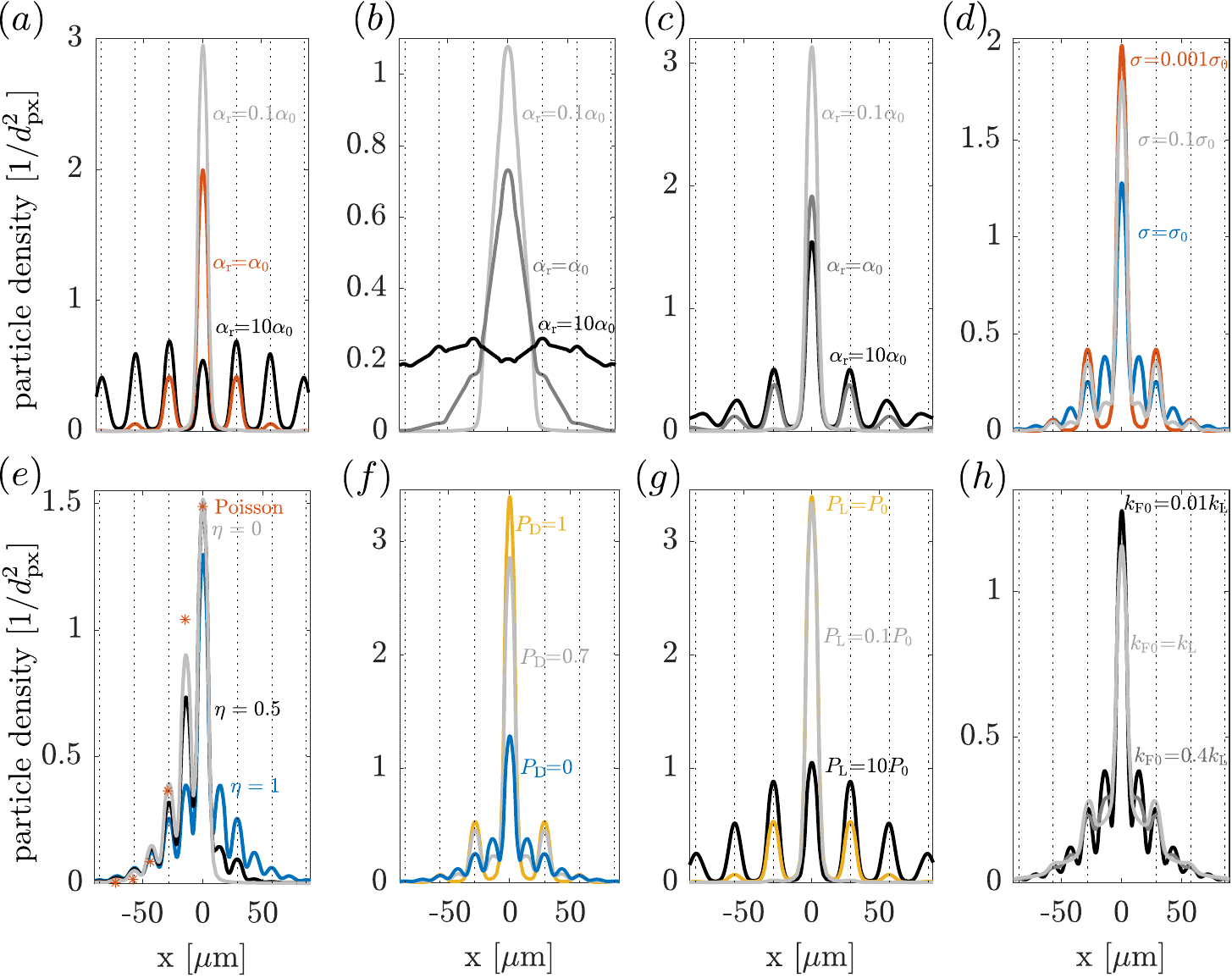}
    \caption{Diffraction patterns illustrating the effect of (a) the polarizability, (b) the polarizability for tripled $x$-slit widths, (c) the polarizability for doubled slit and source widths in $y$-direction, (d) the absorption cross section, (e) the mirror reflectivity, (f) the probability for depletion, (g) the laser power, and (h) fluorescence. We evaluate the interference pattern \eqref{eq:currentDensity} as a function of the horizontal position $x$ at a fixed height $y=-270.30\,{\rm \mu m}$ on the screen. 
    The 2D patterns of corresponding to the red, blue and yellow curve are presented in Fig.~\ref{fig:2Ddiffraction_patterns}. 
    The parameters used in the plots are given in App.\ref{app:simulation_parameters_fig_1}, unless stated otherwise; they are motivated by a real-world experiment \cite{Simonovic2024}.}
    \label{fig:diffraction_patterns}
\end{figure*}

\begin{figure*}[]
    \centering
    \includegraphics[width=0.75\textwidth]{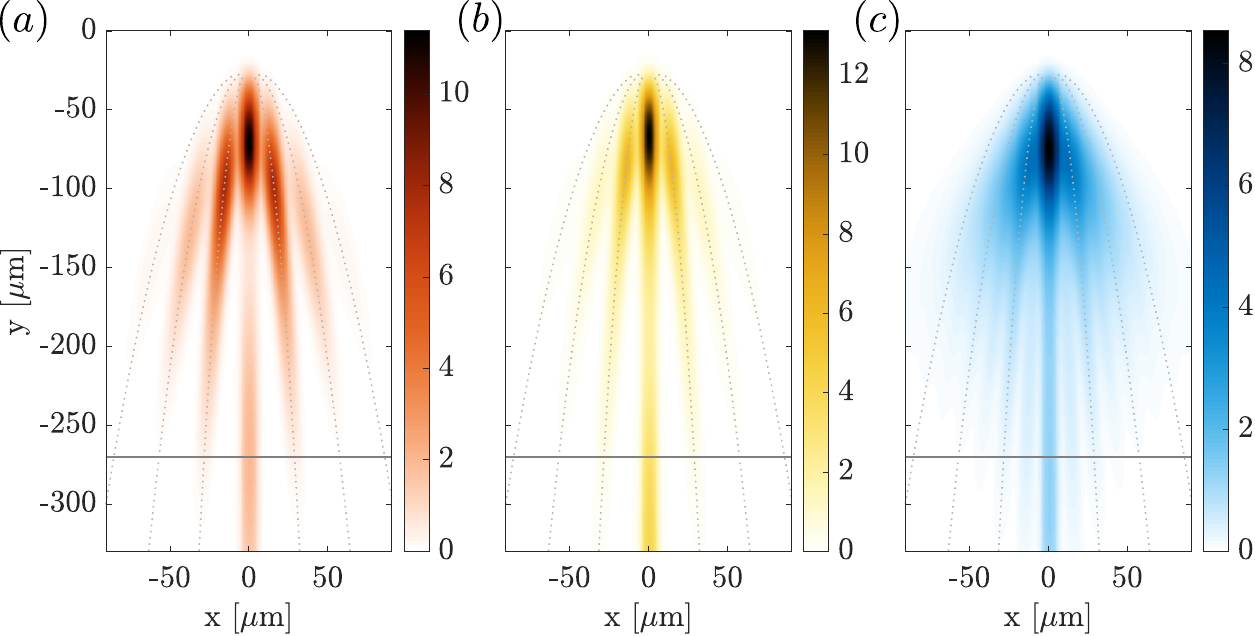}
    \caption{Molecular interference pattern \eqref{eq:currentDensity} for (a) a phase grating, (b) a depletion grating, and (c) an absorption grating. The red, yellow and blue curves in Fig.~\eqref{fig:diffraction_patterns} are obtained by evaluating the densities at the height indicated by the solid gray lines. The dotted lines indicate the expected positions of the diffraction orders for perfect velocity selection, see Sec.~\ref{sec:interference_pattern_polarizability}. Here, the phase grating and the depletion grating give rise to similar density distributions (red and yellow). The distribution of the phase grating (red) stands out through a suppressed zeroth diffraction order, at $y\approx-150\,{\rm \mu m}$. The pattern of the absorption grating (blue) clearly exhibits absorption peaks between the diffraction orders, in contrast to the other two cases. The screen consists of 550$\times$1003 pixels of edge length $d_{\rm px}=0.33\,{\rm \mu m}$. The color indicates the number of molecules per pixel; the total number is $3\times10^{5}$.}
    \label{fig:2Ddiffraction_patterns}
\end{figure*}

In Fig.~\ref{fig:diffraction_patterns}, we display the (time-integrated) current density \eqref{eq:currentDensity} as a function of $x$ at a fixed height on the screen. The figure illustrates the dependence of the interference pattern on different geometrical, molecular and grating parameters, as discussed in the subsequent sections. The red, blue and yellow curves serve as references that connect the panels (a)-(h). We provide plots of the entire 2D screen for these references in Fig.~\ref{fig:2Ddiffraction_patterns}. The particle density on the screen is normalized to a fixed total number of particles.

\subsection{Dependence on the polarizability}\label{sec:interference_pattern_polarizability}
In Fig.~\ref{fig:diffraction_patterns}(a), we vary the polarizability from $\alpha_{\rm r}=0.1\alpha_0$ to $\alpha_{\rm r}=10\alpha_0$. For small polarizabilities (gray line) the interaction with the grating is effectively switched off. In this case, the slit array determines the particle distribution on the screen. Here, we set $\sigma=10^{-3}\sigma_0$ to suppress absorption.

If the polarizability is large enough, the particle is diffracted by the periodic grating potential, giving rise to diffraction peaks (red line). We calculate the peak positions by exploiting that the first $y$-slit is wide enough to be negligible. If the size of $y_{c2}$ and the source are sufficiently small, the height $y$ of a molecule at the screen is related to its momentum $p_z$ through
\begin{equation}\label{eq:pz_of_y}
    p_z(y)=mL_{\rm eff}\left(\blc{\omega}\cdot\blc{e}_x-\sqrt{\frac{g}{L_{\rm eff}}+(\blc{\omega}\cdot\blc{e}_x)^2}\right),
\end{equation}
where the $y$-dependence enters through the effective length 
\begin{equation}
    L_{\rm eff}=\frac{1}{2}\frac{LL_4'(L-L_4')}{L(y-y_{02})+L_4'(y_{\rm 0s}-y)}.
\end{equation}
The ratio of the forward and the grating momentum determines the distances between the diffraction peaks $\Delta x(y)=2\pi\hbar L_4/dp_z(y)$. The $x$-position of the spot where the classical trajectory impinges the screen follows from the arrangement of the two $x$-slits, 
\begin{align}
    x_{\rm class}(y)=&x_{01}+\frac{L_2+L_3+L_4}{L_2}(x_{02}-x_{01})\nonumber\\&
    -\frac{m\blc{\omega}\cdot\blc{e}_y}{p_z(y)}(L_3+L_4)(L_2+L_3+L_4).
\end{align}
The functions $x_{\rm class}(y)$ and $\Delta x(y)$ together define the positions of the dotted lines, which match with the peak positions.

One signature of the phase grating is that low diffraction orders can get suppressed for sufficiently large imprinted phases. This is demonstrated by the black line in Fig.~\ref{fig:diffraction_patterns}(a), where we increased the polarizability by another factor of ten.

\subsection{Effect of greater $x$-slit widths}
In Fig.~\ref{fig:diffraction_patterns}(b), we use the parameters of Fig.~\ref{fig:diffraction_patterns}(a), except for tripled values of the coherence slit widths $x_{\rm c1}$ and $x_{\rm c2}$. In the absence of the phase grating effect (light gray line), the diffraction pattern is clearly broadened due to the wider slits. For larger polarizabilities (dark gray and black line) the diffraction pattern is smeared out, rendering diffraction signatures barely observable.

\subsection{Effect of a greater $y$-slit and source}
In Fig.~\ref{fig:diffraction_patterns}(c), we use the parameters of Fig.~\ref{fig:diffraction_patterns}(a), except for doubled values of $y_{{\rm c}2}$ and $d_{\rm s}$. The increased size of the source and the velocity selection slit imply that the height $y$ on the screen does not unambiguously determine the $z$-momentum through Eq.~\eqref{eq:pz_of_y}. As a consequence, higher diffraction orders smear out and the peak positions do not perfectly coincide with the dotted lines.

\subsection{Dependence on the absorption cross section}
In Fig.~\ref{fig:diffraction_patterns}(d), we vary the absorption cross section from $\sigma=10^{-3}\sigma_0$ to $\sigma=\sigma_0$. For small cross sections, the pattern of a pure phase grating is reproduced (red line). The diffraction peaks are located at the expected positions (dotted lines), compare Fig.~\ref{fig:diffraction_patterns}(a). If the cross section is increased, additional peaks appear right between the existing diffraction peaks, see the gray and blue line. This position results from the fact that only half of the grating momentum is transferred to the molecule during the absorption of a photon.

\subsection{Effect of a finite mirror reflectivity}
In Fig.~\ref{fig:diffraction_patterns}(e), the reflectivity of the mirror is varied. For $\eta=1$, the incident and reflected beam form a standing laser wave. The associated periodic potential landscape gives rise to the blue symmetric diffraction pattern. For $\eta=0$, the grating is given by a running wave, $g(x)=e^{-ik_{\rm L}x}/2$. In this case, there is no periodic potential that diffracts the particle beam. Instead, photon absorption from the running wave produces an asymmetric distribution (gray line), 
which may serve to measure the photon absorption cross section independently from the polarizability.  

A simple expression for the Talbot coefficients is obtained for $\eta=0$ by neglecting fluorescence ($\phi_{\rm F}=0$) and depletion ($P_{\rm D}=0$), and assuming equal polarizabilities and absorption cross sections in the ground and triplet state, see App.~\ref{app:Talbot_coeff_special_case}. This yields
\begin{equation}\label{eq:Talbot_coeff_running_wave}
    B_m(\xi,y)=\delta_{m,0}\exp\left[\frac{1}{4}n_0(y,p_z)\left(e^{i\pi\xi}-1\right)\right].
\end{equation}
The pattern \eqref{eq:currentDensity} contains a $k$-integral that can be decomposed into a convolution of two Fourier transforms. One of them determines the form of the individual peaks of the pattern through the function $\chi_0$ and the position shift due to the Coriolis force through the exponential function. The other Fourier transform defines the relative positions and weights of the peaks,
\begin{align}
    \int_{-\infty}^\infty dk\, e^{ikx}B_0\!\left(\frac{L_4\hbar k}{dp_z}\,,\tilde{y}\right)&=e^{-n}\!\int_{-\infty}^\infty \!dk\, e^{ikx}\exp\!\left(n e^{ikx_0}\right)\nonumber\\
    &=2\pi\sum_{\ell=0}^\infty\frac{n^\ell}{\ell!}e^{-n}\delta(x+\ell x_0)\label{eq:poisson_distribution}.
\end{align}
Here, we inserted Eq.~\eqref{eq:Talbot_coeff_running_wave} and we identified the expected number $n=n_0(\tilde{y},p_z)/4$ of photons that are absorbed during the passage through the grating, see Eq.~\eqref{eq:number_of_absorbed_photons} with $|g(x)|^2=1/4$. The infinite sum follows from expanding $\exp\!\left(n e^{ikx_0}\right)$ into a Taylor series.  The distance $x_0=\pi L_4\hbar/dp_z$ between the peaks results from propagating the momentum kick $\pi\hbar/d$ due to the absorption of a photon over the time $L_4m/p_z$ a molecule needs to travel from the grating to the screen. Since the absorption of several photons occurs independently and at the rate \eqref{eq:absorption_rate2}, they obey a {Poisson distribution}  characterized by the mean number $n$ of absorbed photons.

The Poisson distribution \eqref{eq:poisson_distribution} is displayed by the red stars in Fig.~\ref{fig:diffraction_patterns}(e). We used that for a thin $y$-slit and a small source, the $y$-position at the screen provides information about both $p_z$ and the height $\tilde{y}$, at which the particle traverses the grating. The deviations between the red stars and the maxima of the exact pattern (gray line) are caused by the large size of source and slit. For smaller sizes, the stars and the lines match.

\subsection{Impact of depletion}
In Fig.~\ref{fig:diffraction_patterns}(f), the probability for depletion is varied. If $P_{\rm D}=0$, the absorbed photons cannot cleave or ionize the molecule. In this case, the blue line of Figs.~\ref{fig:diffraction_patterns}(d) and (e) is reproduced. For $P>0$, a fraction of the molecules that absorbed a photon are cleaved or ionized. Since the affected molecules cannot reach the screen the absorption peaks are pronounced less (gray line). For $P_{\rm D}=1$, every molecule that absorbs a photon is removed from the beam and the absorption peaks vanish completely (yellow line). This illustrates that the lack of absorption peaks is no proof for weak absorption.

\subsection{Effect of the laser power on depletion}
In Fig.~\ref{fig:diffraction_patterns}(g), we vary the laser power for the case of efficient depletion, $P_{\rm D}=1$. For low laser power, the expected number of absorbed photons approaches zero. The grating becomes transparent and no diffraction occurs (gray line). For higher laser power, the grating acts like a material grating: Only molecules traversing the grating near the nodes can reach the screen. Consequently, there are only peaks at positions that correspond to the full grating momentum (yellow line). For increasing laser power the absorption rate increases and the effective slit width decreases, implying that higher diffraction orders get populated (black line).

\subsection{Effect of fluorescence}
In Fig.~\ref{fig:diffraction_patterns}(h), we account for fluorescence ($\phi_{\rm F}=1$) and vary the distribution of the fluorescence photons assuming a spectrum of the form
\begin{equation}
    \nu(k)\propto\exp\left[-\frac{(k-k_{\rm F0})^2}{2\sigma_{\rm F}^2}\right]
\end{equation}
with $\sigma_{\rm F}=0.05k_{\rm F0}$. For $k_{\rm F0}\ll k_{\rm L}$ (black line), the emission of the fluorescence photon leads to a negligible momentum kick. The molecular dynamics is then similar to internal conversion, and the pattern coincides with the blue line in the other panels. For photon momenta $\hbar k_{\rm F0}$ comparable to the half grating momentum, the absorption peaks smear out (light and dark gray lines) because of the additional, isotropic momentum kick.

\section{Estimating molecular parameters}\label{sec:measuring_molecular_properties}
We are now in a position to analyze how well molecular parameters can be extracted from  a realistic measurement, where only a finite number of particles is detected and the associated shot noise limits the accuracy of the parameter estimation.

We take the detector to consist of $K$ pixels, each with area $d_{\rm px}^2$. They record a signal $S_k$ proportional to the number $n_k$ of particles detected in the $k$-th pixel, located at position $(x,y)$, during the measurement. The pixels are much smaller than the features of the diffraction pattern \eqref{eq:currentDensity}, so that $p_k=d_{\rm px}^2\rho(x,y)$, with the probability density $\rho(x,y)\propto j(x,y)$ normalized to the screen area $\mathcal{A}_{\rm sc}$,
\begin{equation}
    \int_{\mathcal{A}_{\rm sc}}dxdy\,\rho(x,y)=1.
\end{equation}
In total $N$ independent detections are recorded at the screen, so that the random variables $\hat{S}_k=S_0\hat{n}_k$ are binomially distributed,
\begin{equation}\label{eq:binomialdistribution}
    \hat{n}_k\sim \binom{N}{n_k}p_k^{n_k}(1-p_k)^{N-n_k}.
\end{equation}
The mean signals are thus given by
\begin{subequations}\label{eq:mean_value_and_variance_of_intensity}
\begin{equation}\label{eq:mean_pixel_intensity}
    \bar{S}_{k}=\braket{\hat{S}_k}=S_0Np_k
\end{equation}
and the covariances by
\begin{equation}\label{eq:covariance_pixel_intensity}
    \braket{(\hat{S}_k-\bar{S}_{k})(\hat{S}_\ell-\bar{S}_{\ell})}=\sigma_k^2\delta_{k\ell},
\end{equation}
\end{subequations}
where $\sigma_k^2=S_0^2Np_k(1-p_k)$. 

In the following, we use \eqref{eq:mean_value_and_variance_of_intensity} to calculate the estimation accuracy of molecular parameters for two different fitting procedures.

\subsection{Least-squares fit}\label{sec:normal_fit_general}
A standard least-squares fit returns the set of fit parameters $\alpha=(\alpha_1,...\alpha_{M})$ that minimizes the squared deviation of the measured values $S_k$ from the model function $\bar{S}_{k}(\alpha)$,
\begin{equation}\label{eq:squared_deviation_normal_fit}
    \alpha=\underset{\alpha'}{\rm argmin}\sum_{k=1}^K\left[S_k-\bar{S}_{k}(\alpha')\right]^2.
\end{equation}
To determine the statistics of the fit parameter, as described by the random variable $\hat{\alpha}$, we note that the gradient $\partial/\partial\alpha=(\partial/\partial\alpha_1,...\partial/\partial\alpha_{M})$ vanishes for a given realization $\alpha$,
\begin{equation}\label{eq:gradient_least_squares}
    \sum_{k=1}^K\left[S_k-\bar{S}_{k}(\alpha)\right]\frac{\partial}{\partial\alpha}\bar{S}_{k}(\alpha)=0.
\end{equation}
Assuming the measured intensities $S_k$ are sufficiently close to the mean intensity $\bar{S}_{k}(\alpha_0)$, one can solve \eqref{eq:gradient_least_squares} approximately for $\alpha$. We expand the model function around the actual values $\alpha_0$,
\begin{equation}\label{eq:linearization_intensity}
    \bar{S}_{k}(\alpha)\simeq \bar{S}_{k}(\alpha_0)+(\alpha-\alpha_0)\cdot\frac{\partial}{\partial \alpha_0}\bar{S}_{k}(\alpha_0),
\end{equation}
and insert the linearization into \eqref{eq:gradient_least_squares} to find that the statistics of the fitting parameters are described by
\begin{equation}\label{eq:deviation_least_squares}
\hat{\alpha}=\alpha_0+\mathsf{J}_1^{-1}\hat{\Delta}_1.
\end{equation}
It depends on $\hat{S}_k$ through the tuple
\begin{subequations}
\begin{equation}
    \hat{\Delta}_1=\frac{1}{N^2S_0^2}\sum_{k=1}^K\left[\hat{S}_k-\bar{S}_{k}(\alpha_0)\right]\frac{\partial}{\partial\alpha_0}\bar{S}_{k}(\alpha_0),
\end{equation}
while the symmetric matrix
\begin{equation}
    \mathsf{J}_1=\frac{1}{N^2S_0^2}\sum_{k=1}^K\left[\frac{\partial}{\partial\alpha_0}\bar{S}_{k}(\alpha_0)\right]\otimes \left[\frac{\partial}{\partial\alpha_0}\bar{S}_{k}(\alpha_0)\right]
\end{equation}
\end{subequations}
is independent of the measurement.

The expectation values and covariances of the fit parameters  follow from \eqref{eq:deviation_least_squares} and \eqref{eq:mean_value_and_variance_of_intensity},
\begin{subequations}
\begin{equation}
    \braket{\hat{\alpha}}=\alpha_0,
\end{equation}
\begin{equation}\label{eq:covariance_normal_fit_rho}
\braket{(\hat{\alpha}-\alpha_0)\otimes(\hat{\alpha}-\alpha_0)}=\frac{1}{N}\mathsf{J}_1^{-1}\mathsf{M}\mathsf{J}_1^{-1}.
\end{equation}
\end{subequations}
Here we defined the matrix
\begin{equation}
    \mathsf{M}=\sum_{k=1}^K\frac{\sigma_k^2(\alpha_0)}{N^3S_0^4d_{\rm px}^4}\left[\frac{\partial}{\partial\alpha_0}\bar{S}_{k}(\alpha_0)\right]\otimes \left[\frac{\partial}{\partial\alpha_0}\bar{S}_{k}(\alpha_0)\right].
\end{equation}
Expressing $\sigma_k$ and $\bar{S}_k$ through the probability density $\rho(x,y)$ and performing the limit $\sum_{k=1}^Kd_{\rm px}^2\to \int_{\mathcal{A}_{\rm sc}}dxdy$ yields
\begin{subequations}
\begin{equation}
    \mathsf{J}_1=d_{\rm px}^2\int_{\mathcal{A}_{\rm sc}}dxdy\,\left[\frac{\partial}{\partial\alpha_0}\rho(x,y)\right]\otimes \left[\frac{\partial}{\partial\alpha_0}\rho(x,y)\right]
\end{equation}
and
\begin{align}
     \mathsf{M}=d_{\rm px}^4\int_{\mathcal{A}_{\rm sc}}dxdy\,\rho(x,y)\left[1-d_{\rm px}^2\rho(x,y)\right]\nonumber\\
     \times\left[\frac{\partial}{\partial\alpha_0}\rho(x,y)\right]\otimes \left[\frac{\partial}{\partial\alpha_0}\rho(x,y)\right]
\end{align}
\end{subequations}
Note that $\rho(x,y)$ still depends on the actual values $\alpha_0$. However, if the fit parameters are  approximately known, the covariance \eqref{eq:covariance_normal_fit_rho} serves to quantify the accuracy that can be achieved in an interferometric measurement. In particular, the standard deviation of all the estimated parameters scales as $1/\sqrt{N}$. The matrix $\mathsf{J}_1^{-1}\mathsf{M}\mathsf{J}_1^{-1}$ depends on the specific form of the diffraction pattern and, thus, on all other experimental parameters. The analytical expressions for $\mathsf{J}_1$ and $\mathsf{M}$ can be used to optimize the interferometer for measuring a specific molecular parameter.

\subsection{Iterative fit}\label{sec:iterative_fit_general}
Better accuracy can be achieved if knowledge of the signal shot noise  \eqref{eq:covariance_pixel_intensity} is incorporated in the estimation procedure. One way to do this is to carry out the  fitting procedure in two steps. 

The first step consists of a standard least-squares fit that minimizes \eqref{eq:squared_deviation_normal_fit}, as explained in the previous section. The resulting fit parameters $\alpha_{\rm fit}$ provide an estimate for the shot noise $\sigma^2_k(\alpha_{\rm fit})$ of every pixel. This knowledge is used in the second fit by weighting the intensity of the individual pixels with the estimated variance: a pixel with large uncertainty contributes less than a pixel with small uncertainty. Specifically, the second fit minimizes
\begin{equation}\label{eq:squared_deviation_iterative_fit}
    \alpha=\underset{\alpha'}{\rm argmin}\sum_{k=1}^K\frac{\left[S_k-\bar{S}_{k}(\alpha')\right]^2}{\sigma_k^2(\alpha_{\rm fit})}.
\end{equation}
Finding the fit parameter requires solving 
\begin{equation}\label{eq:gradient_least_squares_iterative}
    \sum_{k=1}^K\frac{\left[S_k-\bar{S}_{k}(\alpha)\right]}{\sigma_k^2(\alpha_{\rm fit})}\frac{\partial}{\partial\alpha}\bar{S}_{k}(\alpha)=0
\end{equation}
for $\alpha$.
If the signal noise is small enough, the estimate form the first fit can be used to approximate $\sigma_k(\alpha_{\rm fit})\simeq\sigma_k(\alpha_0)$. Inserting this into \eqref{eq:gradient_least_squares_iterative}, and solving the equation by the linearization \eqref{eq:linearization_intensity} yields
\begin{equation}\label{eq:deviation_least_squares2}
    \hat{\alpha}=\alpha_0+\mathsf{J}_2^{-1}\hat{\Delta}_2
\end{equation}
with 
\begin{subequations}\label{eq:estimator_iterative_fit}
\begin{equation}
    \hat{\Delta}_2=\sum_{k=1}^K\frac{\left[\hat{S}_k-\bar{S}_{k}(\alpha_0)\right]}{N\sigma_k^2(\alpha_0)}\frac{\partial}{\partial\alpha_0}\bar{S}_{k}(\alpha_0)
\end{equation}
and the symmetric matrix
\begin{equation}
    \mathsf{J}_2=\sum_{k=1}^K\frac{\left[\frac{\partial}{\partial\alpha_0}\bar{S}_{k}(\alpha_0)\right]\otimes \left[\frac{\partial}{\partial\alpha_0}\bar{S}_{k}(\alpha_0)\right]}{N\sigma_k^2(\alpha_0)}.
\end{equation}
\end{subequations}
In the continuum limit it reads as
\begin{equation}\label{eq:fisher_information_div_by_Z}
    \mathsf{J}_2=\int_{\mathcal{A}_{\rm sc}}dxdy\,\frac{\left[\frac{\partial}{\partial\alpha_0}\rho(x,y)\right]\otimes \left[\frac{\partial}{\partial\alpha_0}\rho(x,y)\right]}{\rho(x,y)\left[1-d_{\rm px}^2\rho(x,y)\right]}.
\end{equation}

The expectation values of the fit parameters are again given by $\braket{\hat{\alpha}}=\alpha_0$, while the covariance matrix takes the simple form 
\begin{equation}\label{eq:covariance_iterative_fit_rho}
    \braket{(\hat{\alpha}-\alpha_0)\otimes(\hat{\alpha}-\alpha_0)}=\frac{1}{N}\mathsf{J}_2^{-1}.
\end{equation}
The dependence on the specific form of the diffraction pattern thus differs from the standard fitting procedure. In fact, $N\mathsf{J}_2$ turns into the \emph{Fisher information matrix} \cite{kay1993fundamentals} in the case of small pixels $p_k\ll 1$ and large particle numbers $Np_k\gg 1$. This shows that the iterative fit can reach the highest possible accuracy, as given by the multivariate Cramer-Rao bound \cite{kay1993fundamentals}.

\subsection{Special case: single fit parameter}\label{sec:special_case_one_fit_parameter}
In case of only a single fit parameter, such as the absorption cross section $\sigma$, the variance \eqref{eq:covariance_normal_fit_rho} of the standard fit is given by
\begin{align}\label{eq:variancenormal_one_parameter}
    \langle(\hat{\sigma}&-\sigma_0)^2\rangle=\frac{1}{N}\left\{\int_{\mathcal{A}_{\rm sc}}dx'dy'\,\left[\frac{\partial}{\partial\sigma_0}\rho(x',y')\right]^2\right\}^{-2}\nonumber\\
    &\times\int_{\mathcal{A}_{\rm sc}}dxdy\,\rho(x,y)\left[1-d_{\rm px}^2\rho(x,y)\right]\left[\frac{\partial}{\partial\sigma_0}\rho(x,y)\right]^2,
\end{align}
while the variance \eqref{eq:covariance_iterative_fit_rho} of the iterative fit assumes the form
\begin{equation}\label{eq:varianceiterative_one_parameter}
    \braket{(\hat{\sigma}-\sigma_0)^2}=\frac{1}{N}\left\{\int_{\mathcal{A}_{\rm sc}}dxdy\,\frac{\left[\frac{\partial}{\partial\sigma_0}\rho(x,y)\right]^2}{\rho(x,y)\left[1-d_{\rm px}^2\rho(x,y)\right]}\right\}^{-1}.
\end{equation}

Figure~\ref{fig:setup}(b) shows hundreds of estimates of the absorption cross section (red crosses) as a function of the number $N$ of detected particles. They are based on simulating 2D diffraction patterns (insets) by drawing the number of detected particles in each pixel from the binomial distribution \eqref{eq:binomialdistribution} and performing the iterative fit of Sec.~\ref{sec:iterative_fit_general} by employing two subsequent fit routines.
The dashed lines indicate the expected standard deviation of the estimates, as follows from the analytical formula \eqref{eq:varianceiterative_one_parameter}. The latter  characterizes the distribution of fitted values for $N\gtrsim4\times10^4$. Our calculations imply that $10^6$ particles would suffice to measure the absorption cross section with an accuracy of $\approx0.3\%$ in a realistic experiment, assuming $\sigma=\sigma_{\rm T}$, and all other parameters to be known). 
See App.~\ref{app:simulation_parameters_fig_1} for the simulation parameters. 

\begin{figure}
    \centering
    \includegraphics[width=0.49\textwidth]{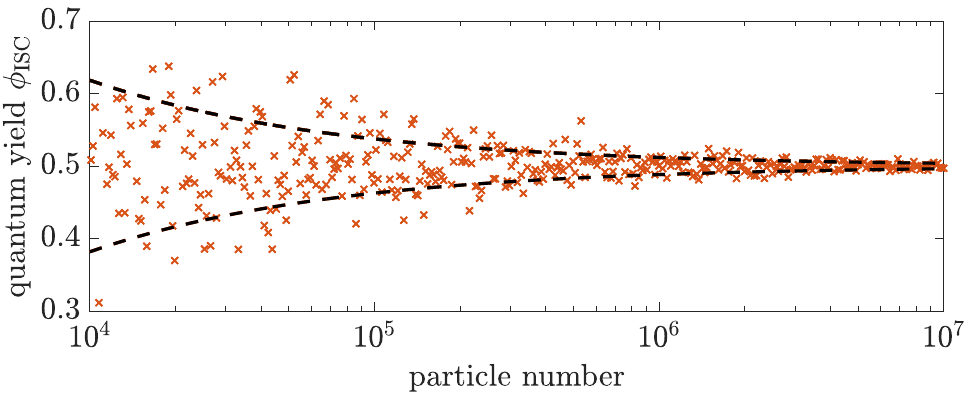}
    \caption{Accuracy of estimating the intersystem-crossing quantum yield by the iterative fitting procedure described in Sec.~\ref{sec:iterative_fit_general}. The standard deviation (dashed line) indicates that a diffraction pattern with $10^6$ particles would suffice to measure the quantum yield $\phi_{\rm ISC}$ with an error of less than $3\%$ for the considered experimental parameters.}
    \label{fig:fit_quantum_yield}
\end{figure}

To demonstrate that matter-wave interferometers can be used to accurately measure quantum yields in vacuum,   we analyze
in Fig.~\ref{fig:fit_quantum_yield} the intersystem-crossing quantum yield $\phi_{\rm ISC}$. The measurements are again simulated with the parameters given in App.~\ref{app:simulation_parameters_fig_1}, except for $\alpha_{\rm r}=\alpha_{\rm r0}$, $\alpha_{\rm rT}=2\alpha_{\rm r0}$, $\phi_{\rm ISC}=0.5$, and $\phi_{\rm IC}=0.5$. Note that the model function $\bar{S}_k(\phi_{\rm ISC})$ accounts for the fact that a varying intersystem-crossing quantum yield $\phi_{\rm ISC}$ also changes the internal-conversion quantum yield $\phi_{\rm IC}=1-\phi_{\rm ISC}$. We find that an accuracy of better than 3\% can be reached by detecting $10^6$ particles.

\subsection{Special case: two fit parameters}

We next illustrate the case of measuring two molecular parameters simultaneously, specifically the absorption cross section $\sigma$ and the polarizability $\alpha_{\rm r}$.
For the standard least-squares fit the variances in \eqref{eq:covariance_normal_fit_rho} assume the form
\begin{subequations}\label{eq:variances_two_parameters_normal_fit}
\begin{align}\label{eq:variance_sigma_two_parameters_normal_fit}
\langle(&\hat{\sigma}-\sigma_0)^2\rangle=\frac{1}{N}\frac{1}{\left[b_{\sigma\sigma}b_{\alpha\alpha}-b_{\sigma\alpha}^2\right]^2}\int_{\mathcal{A}_{\rm sc}}dxdy\,\rho(x,y)\nonumber\\
&\times\left[1-d_{\rm px}^2\rho(x,y)\right]\left[b_{\alpha\alpha}\frac{\partial}{\partial\sigma_0}\rho(x,y)-b_{\sigma\alpha}\frac{\partial}{\partial\alpha_{\rm r0}}\rho(x,y)\right]^2
\end{align}
and
\begin{align}\label{eq:variance_alpha_two_parameters_normal_fit}
\langle(&\hat{\alpha}_{\rm r}-\alpha_{\rm r0})^2\rangle=\frac{1}{N}\frac{1}{\left[b_{\sigma\sigma}b_{\alpha\alpha}-b_{\sigma\alpha}^2\right]^2}\int_{\mathcal{A}_{\rm sc}}dxdy\,\rho(x,y)\nonumber\\
&\times\left[1-d_{\rm px}^2\rho(x,y)\right]\left[b_{\sigma\sigma}\frac{\partial}{\partial\alpha_{\rm r0}}\rho(x,y)-b_{\sigma\alpha}\frac{\partial}{\partial\sigma_0}\rho(x,y)\right]^2,
\end{align}
\end{subequations}
where
\begin{subequations}
\begin{align}
    b_{\sigma\sigma}&=\int_{\mathcal{A}_{\rm sc}}dxdy\,\left[\frac{\partial}{\partial\sigma_0}\rho(x,y)\right]^2,\\
    b_{\alpha\alpha}&=\int_{\mathcal{A}_{\rm sc}}dxdy\,\left[\frac{\partial}{\partial\alpha_{\rm r0}}\rho(x,y)\right]^2,\\
    b_{\sigma\alpha}&=\int_{\mathcal{A}_{\rm sc}}dxdy\,\left[\frac{\partial}{\partial\sigma_0}\rho(x,y)\right]\left[\frac{\partial}{\partial\alpha_{\rm r0}}\rho(x,y)\right].
\end{align}
\end{subequations}

For the iterative fit with two fit parameters, the standard deviation follows from \eqref{eq:covariance_iterative_fit_rho},
\begin{subequations}\label{eq:variances_two_parameters_iterative_fit}
\begin{align}\label{eq:variance_sigma_two_parameters_iterative_fit}
\langle(&\hat{\sigma}-\sigma_0)^2\rangle=\frac{1}{N}\frac{c_{\alpha\alpha}}{c_{\sigma\sigma}c_{\alpha\alpha}-c_{\sigma\alpha}^2}
\end{align}
and
\begin{align}\label{eq:variance_alpha_two_parameters_iterative_fit}
\langle(&\hat{\alpha}_{\rm r}-\alpha_{\rm r0})^2\rangle=\frac{1}{N}\frac{c_{\sigma\sigma}}{c_{\sigma\sigma}c_{\alpha\alpha}-c_{\sigma\alpha}^2}
\end{align}
\end{subequations}
with
\begin{subequations}
\begin{align}
    c_{\sigma\sigma}&=\int_{\mathcal{A}_{\rm sc}}dxdy\,\frac{\left[\frac{\partial}{\partial\sigma_0}\rho(x,y)\right]^2}{\rho(x,y)\left[1-d_{\rm px}^2\rho(x,y)\right]},\\
    c_{\alpha\alpha}&=\int_{\mathcal{A}_{\rm sc}}dxdy\,\frac{\left[\frac{\partial}{\partial\alpha_{\rm r0}}\rho(x,y)\right]^2}{\rho(x,y)\left[1-d_{\rm px}^2\rho(x,y)\right]},\\
    c_{\sigma\alpha}&=\int_{\mathcal{A}_{\rm sc}}dxdy\,\frac{\left[\frac{\partial}{\partial\sigma_0}\rho(x,y)\right]\left[\frac{\partial}{\partial\alpha_{\rm r0}}\rho(x,y)\right]}{\rho(x,y)\left[1-d_{\rm px}^2\rho(x,y)\right]}.
\end{align}
\end{subequations}

\begin{figure}
    \centering
    \includegraphics[width=0.49\textwidth]{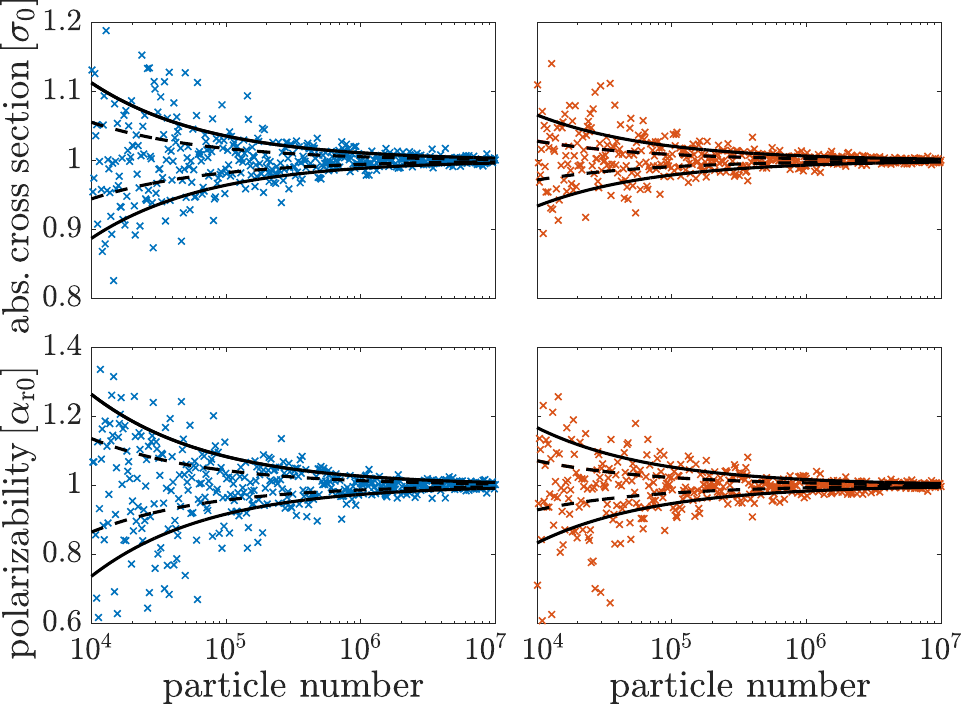}
    \caption{Accuracy of the normal (blue) and the iterative (red) fit as described in Secs.~\ref{sec:normal_fit_general} and \ref{sec:iterative_fit_general}. The solid lines represent the analytically expected standard deviation, see Eqs.~\eqref{eq:variances_two_parameters_normal_fit} and Eqs.~\eqref{eq:variances_two_parameters_iterative_fit}. The dashed lines give the expected standard deviation if only a single parameter was fitted, see Eq.~\eqref{eq:variancenormal_one_parameter} and Eq.~\eqref{eq:varianceiterative_one_parameter}. By detecting  $10^6$ particles on the screen, both the absorption cross section and the polarizability could be obtained from an iterative fit with an accuracy of better than $2\%$. (Here we assume  $\sigma=\sigma_{\rm T}$ and $\alpha_{\rm r}=\alpha_{\rm rT}$.)}
    \label{fig:fit_statistics}
\end{figure}

Figure~\ref{fig:fit_statistics} shows the expected results of a simultaneous measurement of the absorption cross section and the polarizability for standard (blue, left) and iterative (red, right) fitting procedures. The solid lines give the for the expected standard deviations of the absorption cross section and the polarizability according to the analytic expressions \eqref{eq:variances_two_parameters_normal_fit}  and \eqref{eq:variances_two_parameters_iterative_fit}. Like in  Fig.~\ref{fig:setup}(b), we confirm that the presented formulas provide realistic estimates for the measurement accuracy by simulating detections on the screen and estimating the two parameters through unbiased fit routines that minimize \eqref{eq:squared_deviation_normal_fit} (blue crosses) and \eqref{eq:squared_deviation_iterative_fit} (red crosses), respectively.

The numerous fits and the predicted accuracies demonstrate the advantages of the iterative procedure, which decreases the error by roughly a factor of two. A similar gain in accuracy is achieved if the number of fit parameters is reduced to one, as indicated by the dashed lines in Fig.~\ref{fig:fit_statistics}. They give the standard deviation if only $\sigma$ or only $\alpha_{\rm r}$ were unknown, see Sec.~\ref{sec:special_case_one_fit_parameter}. 
To account for possible uncertainties of parameters that are not included in the fitting procedure, one can use Eqs.~\eqref{eq:deviation_least_squares} and \eqref{eq:deviation_least_squares2} 
for error propagation. \tcb{To calculate the accuracy in simultaneous fits with three or more fit parameters, we refer the reader to the general formulas \eqref{eq:covariance_normal_fit_rho} and \eqref{eq:covariance_iterative_fit_rho}.}

\section{Conclusions and Outlook}\label{sec:conclusions}

The results  presented in this article show that photophysical molecular properties can be measured in vacuum using matter-wave diffraction off a single light grating.  The explicit formula for the 2D interference pattern \eqref{eq:currentDensity} accounts for all relevant effects and elements in a realistic setup: from finite-sized source and slits determining the longitudinal and transverse coherence, to gravitational velocity selection, and deflection due to the Coriolis force. Besides accounting for the Gaussian envelope of the laser grating and the finite reflectivity of the grating mirror, our formula incorporates the full range of photophysical processes that determine the interaction of a molecule with a laser wave. In addition to inducing a dipole, this includes  internal conversion, intersystem crossing, fluorescence, ionization, and cleavage
following the absorption of a real photon.

Using this theory, the accuracy of interferometric measurements of molecular parameters could be estimated from realistic, shot-noise-limited diffraction patterns. We find that quantities such as polarizabilities, absorption cross sections, and quantum yields can be determined with uncertainties on the single percent level in a generic state-of-the-art setup.
To improve on that, the formulas in Sec.~\ref{sec:measuring_molecular_properties} can be used to optimize the interferometric setup for measuring specific photophysical parameters with the highest possible accuracy. 
In practice, the accuracy will be determined by the shot noise due to the limited  particle beam brilliance and detection efficiency, while the precision is determined by the uncertainty of the experimental parameters. Both can be assessed by using Eqs.~\eqref{eq:deviation_least_squares} and \eqref{eq:deviation_least_squares2} for the statistics of the extracted parameters using a standard or an iterative fitting procedure.

\section*{Acknowledgements}
We acknowledge support  by the German Aerospace Center (DLR) within project 50WM2264 with funds provided by the Federal Ministry for Economic Affairs and Climate Action (BMWK), by the Austrian Science Funds (FWF) within project DCAFM Hidhys and P32543-N, by the European Commission within project 860713, by the Gordon \& Betty Moore Foundation within project 10771, and by the Open Access Publication Fund of the University of Duisburg-Essen.

\appendix
\section{Wigner function and characteristic function}\label{app:basic_trafos}
When describing the quantum dynamics of a molecule traveling through the interferometer, it is safe to assume that the $p_z$-momentum of the molecule is not affected by the diffractive elements. The $z$-coordinate then parametrizes time, $t=mz/p_z$. Furthermore, we assume that diffraction in $y$-direction is negligible and quantum signatures appear only in $x$-direction. To describe the latter, we employ the {Wigner function}
\begin{equation}\label{eq:definition_wigner_function}
    w(x,p_x)=\frac{1}{2\pi\hbar} \int ds\,e^{isp_x/\hbar}\bra{x-\frac{s}{2}}\uprho\ket{x+\frac{s}{2}},
\end{equation}
which can be calculated from the statistical operator $\uprho$. It is the natural quantum generalization of the classical phase-space distribution.

The time-evolution in the interferometer can be divided into periods of uniformly accelerated motion and instantaneous transformations at the diffractive elements. The evolution of an initial Wigner function $w(x,p_x)$ over the time $t$ in presence of an acceleration $a$ reads \cite{nimmrichter2014springer}
\begin{equation}\label{eq:definition_free_prop_Wigner}
    w'(x,p_x)=w\!\left(x-\frac{p_xt}{m}+\frac{at^2}{2},p_x-mat\right).
\end{equation}
It is indistinguishable from the classical evolution in phase space. Note the positive sign of the term $at^2/2$. When passing an $x$-slit or a grating, the Wigner function transforms as
\begin{equation}\label{eq:wigner_funct_transform_grating_slit}
    w'(x,p_x)=\int_{-\infty}^\infty dp\, T(x,p_x-p)w(x,p),
\end{equation}
disregarding normalization. The convolution kernel of a collimation slit of width $x_{\rm c}$ at the position $x_0$ is given by
\begin{align}\label{eq:slit_trafo_quantum}
    T(x,p_x)=&\Theta\!\left(\frac{x_{\rm c}}{2}-|x-x_{0}|\right)\frac{x_{\rm c}-2|x-x_{0}|}{\pi\hbar}\nonumber\\
    &\times{\rm sinc}\!\left(\frac{x_{\rm c}-2|x-x_{0}|}{\hbar}p_x\right).
\end{align}
For $\hbar\to 0$, it yields the classical convolution kernel 
\begin{equation}\label{eq:slit_trafo_classical}
    T(x,p_x)=\Theta\!\left(\frac{x_{\rm c}}{2}-|x-x_{0}|\right)\delta(p_x).
\end{equation}
The kernel of a grating is determined by the Talbot coefficients \cite{hornberger2011,nimmrichter2014springer},
\begin{equation}\label{eq:fourier_series_T_Talbot_coeff}
    T(x,p_x)=\frac{1}{2\pi\hbar}\sum_{n=-\infty}^\infty e^{2\pi inx/d}\int ds\,e^{ip_xs/\hbar}B_n\!\left(\frac{s}{d}\right).
\end{equation}
It is convenient to use the {characteristic function} \cite{bateman2014}
\begin{equation}\label{eq:definition_characteristic_funct}
    \chi(s,q)=\int dxdp_x\,e^{i(qx-sp_x)/\hbar}w(x,p_x)
\end{equation}
as an alternative to the Wigner function. Its propagation in presence of a constant acceleration obeys
\begin{equation}\label{eq:propagation_char_funct}
    \chi'(s,q)=\exp\!\left[\frac{i}{\hbar}\left(\frac{at^2q}{2}-mats\right)\right]\chi\!\left(s-\frac{qt}{m},q\right),
\end{equation}
in accordance with Eq.~\eqref{eq:definition_free_prop_Wigner}. The major benefit of the characteristic function is that its transformation at a grating takes a simple form,
\begin{equation}\label{eq:grating_trafo_characteristic_funct}
    \chi'(s,q)=\sum_{n=-\infty}^\infty B_n\!\left(\frac{s}{d}\right)\chi\!\left(s,q+\frac{2\pi\hbar}{d}n\right).
\end{equation}

\section{Derivation of the Talbot coefficients}\label{app:derivation_Talbot_coeff}
\subsection{Master equation}\label{app:derivation_Talbot_master_eq}
To derive the Talbot coefficients, the dynamics of the molecule traversing the laser grating have to be solved. The molecule-grating interaction involves the processes described in Sec.~\ref{sec:particle-grating_interaction} and displayed in Fig.~\ref{fig:internal_processes}. We account for all of them by considering the master equation
\begin{align}\label{app:general_master_equation}
\partial_t\uprho=&\frac{1}{i\hbar}[{\rm H},\uprho]+\sum_{j\in\mathcal{M}}\left({\rm L}_j\uprho {\rm L}_j^\dagger-\frac{1}{2}\left\{{\rm L}_j^\dagger {\rm L}_j,\uprho\right\}\right)\nonumber\\
&+\int \frac{d^3k}{4\pi k^2}f_k(k)\left({\rm L}_{\rm F}\uprho {\rm L}_{\rm F}^\dagger-\frac{1}{2}\left\{{\rm L}_{\rm F}^\dagger {\rm L}_{\rm F},\uprho\right\}\right),
\end{align}
in analogy to the treatment in \cite{walterDiss}. The coherent interaction due to the polarizability of the molecule is described in  the eikonal approximation by the generator
\begin{align}
{\rm H}=V({ x},p_zt/m)\ket{G}\bra{G}+V_{\rm T}({ x},p_zt/m)\ket{T}\bra{T},
\end{align}
where the position operator ${\rm x}$ appears in the potential \eqref{eq:potential_definition}. The internal molecular transitions and the associated momentum kicks are described by a set of Lindblad operators,
\begin{subequations}
\begin{align}
{\rm L}_{\rm IC}=&\sqrt{(1-P_{\rm D})\gamma_{\rm max}(t)\phi_{\rm IC}}\,g({\rm x})\ket{G}\bra{G},\\
{\rm L}_{\rm F}=&\sqrt{(1-P_{\rm D})\gamma_{\rm max}(t)\phi_{\rm F}}\,g({\rm x})e^{-i\blc{k}\cdot\blc{e}_x {\rm x}}\ket{G}\bra{G},\\
{\rm L}_{\rm ISC}=&\sqrt{(1-P_{\rm D})\gamma_{\rm max}(t)\phi_{\rm ISC}}\,g({\rm x})\ket{T}\bra{G},\\
{\rm L}_{\rm ICT}=&\sqrt{(1-P_{\rm D})\gamma_{\rm max,T}(t)}\,g({\rm x})\ket{T}\bra{T},\\
{\rm L}_{\rm DG}=&\sqrt{P_{\rm D}\gamma_{\rm max}(t)}\,g({\rm x})\ket{D_G}\bra{G},\\
{\rm L}_{\rm DT}=&\sqrt{P_{\rm D}\gamma_{\rm max,T}(t)}\,g({\rm x})\ket{D_T}\bra{T}.
\end{align}
\end{subequations}
The different processes are indicated by the arrows in Fig.~\ref{fig:internal_processes} and labeled by $\mathcal{M}=\{{\rm IC,ISC,ICT,DG,DT}\}$. The Lindblad operators depend on the peak value
\begin{align}\label{eq:maximum_value_absorption_rates}
\gamma_{\rm max}(t)&=\frac{8P_{\rm L}\sigma}{\pi \hbar \omega_{\rm L}w_yw_z}
    f^2(y,p_zt/m),\\
\gamma_{\rm max,T}(t)&=\frac{8P_{\rm L}\sigma_{\rm T}}{\pi \hbar \omega_{\rm L}w_yw_z}
    f^2(y,p_zt/m),
\end{align}
of the absorption rate \eqref{eq:absorption_rate2}.

\begin{widetext}

Differential equations for the matrix elements $\rho_{\rm G}(x,x',t)=\bra{x,G}\uprho\ket{x',G}$ and $\rho_{\rm T}(x,x',t)=\bra{x,T}\uprho\ket{x',T}$ follow from the master equation \eqref{app:general_master_equation}. Here, $\ket{x,G}\equiv\ket{x}\ket{G}$ and $\ket{x,T}\equiv\ket{x}\ket{T}$ depend on the eigenstates $\ket{x}$ of the position operator ${\rm x}$. If the particle detector at the screen is insensitive to the internal molecular state the measurement will only yield information on the unconditional state $\rho_{\rm un}(x,x',t)=\rho_{\rm G}(x,x',t)+\rho_{\rm T}(x,x',t)$. Since molecules in a depletion state do not reach the screen, these states are disregarded. 
We find a coupled system of differential equations,
\begin{subequations}\label{eq:DGLn_uncorrelated_state}
\begin{align}
\partial_t\rho_{\rm un}(x,x',t)=&\left[\tilde{D}_{\rm T}(t)+\tilde{N}_{\rm T}(t)\right]\rho_{\rm un}(x,x',t)+\left\{\tilde{D}(t)-\tilde{D}_{\rm T}(t)+\tilde{N}(t)[\phi_{\rm IC}+\phi_{F}\varphi(x-x')+\phi_{\rm ISC}]-\tilde{N}_{\rm T}(t)\right\}\rho_{\rm G}(x,x',t),\label{eq:diff_eq_uncond1}\\
\partial_t\rho_{\rm G}(x,x',t)=&\left\{\tilde{D}(t)+\tilde{N}(t)[\phi_{\rm IC}+\phi_{\rm F}\varphi(x-x')]\right\}\rho_{\rm G}(x,x',t)\label{eq:diff_eq_uncond2}
\end{align}
\end{subequations}
The time-dependent functions in Eqs.~\eqref{eq:DGLn_uncorrelated_state} are related to the definitions \eqref{eq:definitions_DN} from the main text,
\begin{align}
    \tilde{D}(t)=&D\sqrt{\frac{2}{\pi}}\frac{p_z}{mw_z}\exp\!\left(-2\frac{p_z^2t^2}{m^2w_z^2}\right),
\end{align}
and similar for $\tilde{D}_{\rm T}$, $\tilde{N}$ and $\tilde{N}_{\rm T}$.
Note that one could also start out with a level scheme that accounts for the fact that the vibrational state of the molecule changes after every internal conversion or fluorescence. However, the resulting system of differential equations for the unconditional state with respect to the electronic and vibrational degrees of freedom would be equal to \eqref{eq:DGLn_uncorrelated_state}.

\subsection{Solving the master equation}
Before interacting  with the laser grating the molecule is taken to be in its electronic ground state $\rho_{\rm G}(x,x',-\infty)=\tilde{\rho}(x,x')$. In this case, Eq.~\eqref{eq:diff_eq_uncond2} has the formal solution
\begin{align}\label{eq:solution_rho_G}
\rho_{\rm G}(x,x',t)=&\tilde{\rho}(x,x')\exp\!\left(\int_{-\infty}^t\,d\tau\left\{\tilde{D}(\tau)+\tilde{N}(\tau)[\phi_{\rm IC}+\phi_{\rm F}\varphi(x-x')]\right\}\right).
\end{align}
We insert \eqref{eq:solution_rho_G} into \eqref{eq:diff_eq_uncond1} and use that initially $\rho_{\rm un}(x,x',-\infty)=\tilde{\rho}(x,x')$ to find
\begin{align}
\rho_{\rm un}(x,x',t)=&\tilde{\rho}(x,x')\exp\left\{\int_{-\infty}^td\tau''[\tilde{D}_{\rm T}(\tau'')+\tilde{N}_{\rm T}(\tau'')]\right\}\left(1+\int_{-\infty}^td\tau'\{\tilde{D}(\tau')-\tilde{D}_{\rm T}(\tau')+\tilde{N}(\tau')[\phi_{\rm ISC}+\phi_{\rm IC}\right.\nonumber\\
&\left.+\phi_{\rm F}\varphi(x-x')]-\tilde{N}_{\rm T}(\tau')\}\exp\!\left[\int_{-\infty}^{\tau'}d\tau\{\tilde{D}(\tau)-\tilde{D}_{\rm T}(\tau)+\tilde{N}(\tau)[\phi_{\rm IC}+\phi_{\rm F}\varphi(x-x')]-\tilde{N}_{\rm T}(\tau)\}\right]\right).
\end{align}
In a distance several laser widths behind the grating, the state can be approximated by 
\begin{align}\label{eq:rho_infty_gaussian_ints}
  \rho_{\rm un}(x,x',\infty)=&\tilde{\rho}(x,x')\left\{1+\nu_1\int_{-\infty}^\infty d\tau'\,\exp\!\left(-2\frac{p_z^2{\tau'}^2}{m^2w_z^2}\right)\exp\!\left[\nu_2\int_{-\infty}^{\tau'}d\tau \exp\!\left(-2\frac{p_z^2\tau^2}{m^2w_z^2}\right)\right]\right\}\nonumber\\
  &\times \exp\!\left[\nu_3\int_{-\infty}^\infty d\tau'' \exp\!\left(-2\frac{p_z^2{\tau''}^2}{m^2w_z^2}\right)\right].
\end{align}
Here, we defined the frequencies
\begin{subequations}
\begin{align}
    \nu_1=&\sqrt{\frac{2}{\pi}}\frac{p_z}{mw_z}\{D-D_{\rm T}+N[\phi_{\rm IC}+\phi_{\rm F}\varphi(x-x')+\phi_{\rm ISC}]-N_{\rm T}\}\\
    \nu_2=&\sqrt{\frac{2}{\pi}}\frac{p_z}{mw_z}\{D-D_{\rm T}+N[\phi_{\rm IC}+\phi_{\rm F}\varphi(x-x')]-N_{\rm T}\},\\
    \nu_3=&\sqrt{\frac{2}{\pi}}\frac{p_z}{mw_z}(D_{\rm T}+N_{\rm T}).
\end{align}
\end{subequations}
We use that
\begin{align}
\int_{-\infty}^{\infty}dz'\,e^{-2z'^2}\exp\!\left[\vartheta\int_{-\infty}^{z'}dz \,e^{-2z^2}\right]=\frac{1}{\vartheta}\left(e^{\vartheta\sqrt{\pi/2}}-1\right)
\end{align}
to evaluate the Gaussian time integrals in Eq.~\eqref{eq:rho_infty_gaussian_ints}, 
\begin{align}\label{eq:unconditioned_rho_final_result}
    \rho_{\rm un}(x,x',\infty)=&\left\{1+\frac{\nu_1}{\nu_2}\left[\exp\left(\sqrt{\frac{\pi}{2}}\frac{w_zm}{p_z}\nu_2\right)-1\right]\right\}
    \exp\left(\sqrt{\frac{\pi}{2}}\frac{w_zm}{p_z}\nu_3\right)\tilde{\rho}(x,x').
\end{align}

\subsection{Talbot coefficients}\label{app:talbot_coefficients}
At the grating, the Wigner function transforms as in \eqref{eq:wigner_funct_transform_grating_slit}, and the state after the grating is of the form $\rho_{\rm un}(x,x',\infty)=F(x,x')\tilde{\rho}(x,x')$. This corresponds to a transmission kernel
\begin{equation}\label{eq:transmission_kernel_F}
    T(x,p)=\frac{1}{2\pi\hbar}\int ds\,e^{isp/\hbar}F\!\left(x-\frac{s}{2},x+\frac{s}{2}\right),
\end{equation}
as follows from the definition \eqref{eq:definition_wigner_function} of the Wigner function. Since the $F$ in Eq.~\eqref{eq:transmission_kernel_F} is $d$-periodic in $x$, it can be expanded as a Fourier series, see Eq.~\eqref{eq:fourier_series_T_Talbot_coeff}. The {Talbot coefficients}
\begin{equation}\label{eq:FT_Talbot_coefficients_appendix}
B_n(\xi,y)=\frac{1}{d}\int_{-d/2}^{d/2}dx\,e^{-2\pi inx/d}F\!\left(x-\xi\frac{d}{2},x+\xi\frac{d}{2}\right)
\end{equation}
act as the Fourier coefficients. The state right behind the grating \eqref{eq:unconditioned_rho_final_result} implies that
\begin{align}
    F(x,x')=&\left\{1+\frac{\nu_1}{\nu_2}\left[\exp\left(\sqrt{\frac{\pi}{2}}\frac{w_zm}{p_z}\nu_2\right)-1\right]\right\}
    \exp\left(\sqrt{\frac{\pi}{2}}\frac{w_zm}{p_z}\nu_3\right),
\end{align}
which immediately yields Eq.~\eqref{eq:F_maintext}. 

The function \eqref{eq:F_maintext} thus describes the interaction of a molecule with a laser grating. It accounts for the imprinted phase, photon absorption, state-dependent polarizabilities and absorption cross sections, internal conversion, intersystem crossing, fluorescence, photo-cleavage, and ionization as well as non-ideal reflection at the grating mirror. Remarkably, for $\eta=1$ and $P_{\rm D}=0$ we reproduce the Talbot coefficients obtained in \cite{walterDiss}, although the calculation in \cite{walterDiss} assumes an effective rectangular laser profile.

\subsection{Special case}\label{app:Talbot_coeff_special_case}
The Fourier transform in Eq.~\eqref{eq:FT_Talbot_coefficients_appendix} can be evaluated analytically for the special case that there is no fluorescence and the triplet state has the same optical response as the ground state, $\alpha_{\rm r, T}=\alpha_{\rm r}$, $\sigma_{\rm T}=\sigma$ and $\phi_{\rm F}=0$. In this case the imprinted phase and the photon number coincide for the singlet and triplet state, $\phi_{0,{\rm T}}=\phi_0$ and $n_{0,{\rm T}}=n_0$, so that $D_{\rm T}=D$ and $N_{\rm T}=N$. Equation~\eqref{eq:F_maintext} then takes the simple form $F(x,x')=\exp({D+N})$, explicitly
\begin{align}
    F(x,x')=
    \exp\left[i\left(|g(x)|^2-|g(x')|^2\right)\phi_0(y,p_z)
    -\frac{1}{2}\left(|g(x)|^2+|g(x')|^2\right)n_0(y,p_z)
    +(1-P_{\rm D})g^*(x')g(x)n_0(y,p_z)
    \right].\label{eq:F_special_case}
\end{align}
Inserting \eqref{eq:definition_gx} and \eqref{eq:F_special_case} into \eqref{eq:FT_Talbot_coefficients_appendix} yields
\begin{align}\label{eq:Talbot_coeff_zeta_definition}
    B_n(\xi,y)=&\frac{1}{2\pi}\int_{-\pi}^\pi dx'\,\exp\!\left[-inx'+i\zeta_{\rm coh}(\xi)\sin(x')
    +\zeta_{\rm abs}(\xi)\cos(x')-\kappa(\xi)\right].
\end{align}
Here, we defined
\begin{align}
    \zeta_{\rm coh}(\xi)=&\eta \sin(\pi\xi) \phi_0(y,p_z),\\
    \zeta_{\rm abs}(\xi)=&\eta \left[\sin^2\!\left(\frac{\pi\xi}{2}\right)-\frac{P_{\rm D}}{2}\right]n_0(y,p_z),\\
    \kappa(\xi)=&\frac{\eta^2+1}{4}n_0(y,p_z)-(1-P_{\rm D})\left[\frac{\eta^2+1}{4}\cos(\pi\xi)
    -i\frac{\eta^2-1}{4}\sin(\pi\xi)\right]n_0(y,p_z).
\end{align}
The integral in Eq.~\eqref{eq:Talbot_coeff_zeta_definition} can be evaluated \cite{hornberger2011,bateman2014,walterDiss,nimmrichter2014springer}, resulting in the Talbot coefficients
\begin{align}\label{eq:special_Talbot_coefficients_results}
    B_n(\xi,y)=&e^{-\kappa}J_n\!\left[{\rm sgn(\zeta_{\rm coh}-\zeta_{\rm abs})}\sqrt{\zeta^2_{\rm coh}-\zeta^2_{\rm abs}}\right]
    \left({\frac{\zeta_{\rm coh}+\zeta_{\rm abs}}{\zeta_{\rm coh}-\zeta_{\rm abs}}}\right)^{n/2}.
\end{align}
In an extension to previous results \cite{bateman2014,nimmrichter2014springer,cotter2015,walter2016multiphoton,walterDiss}, the  Talbot coefficients \eqref{eq:special_Talbot_coefficients_results} account for depletion with finite probability $P_{\rm D}<1$, for non-ideal reflection at the grating mirror, and for the Gaussian shape of the laser wave. For $\eta=1$ and $P_{\rm D}=1$, Eq.~\eqref{eq:special_Talbot_coefficients_results} reproduces the Talbot coefficients of a single-photon ionization grating \cite{nimmrichter2011}. For $\eta=1$ and $P_{\rm D}=0$, it reproduces the Talbot coefficients for multiphoton absorption \cite{bateman2014,nimmrichter2014springer,walter2016multiphoton}.

\section{Derivation of the interference pattern}\label{app:calculating_pattern}
In this section, we calculate the current density (\ref{eq:currentDensity}) of a particle beam passing through the interferometer shown in Fig.~\ref{fig:setup}. At the source, the the Wigner function is indistinguishable from  the classical distribution function,
\begin{align}\label{eq:initial_state_app}
    w_{\rm s}(x,y,\blc{p})=& \frac{1}{d_{\rm s}^2}\Theta\!\left(\frac{d_{\rm s}}{2}-|x-x_{0{\rm s}}|\right)\Theta\!\left(\frac{d_{\rm s}}{2}-|y-y_{0{\rm s}}|\right)\mu_x(p_x)\mu_y(p_y)\mu_z(p_z).
\end{align}
In the following, the Wigner function for the particle state is propagated through the interferometer by using the transformations introduced in App.~\ref{app:basic_trafos}. We 
assume that diffraction into the direction parallel to the grating slits plays no role, as is the case in many interferometers, so that the $y$-dynamics can be treated classically.

In a first step, the initial state \eqref{eq:initial_state_app} is propagated over a distance $L_1$ from the source to the first $x$-slit, yielding
\begin{align}
    w_{\rm L_1}(x,y,\blc{p})= &\frac{1}{d_{\rm s}^2}\Theta\!\left(\frac{d_{\rm s}}{2}-\left|x-x_{0{\rm s}}-L_1\frac{p_x}{p_z}+\frac{a_{{\rm c},x}L_1^2m^2}{2p_z^2}\right|\right)\Theta\!\left(\frac{d_{\rm s}}{2}-\left|y-y_{0{\rm s}}-L_1\frac{p_y}{p_z}+\frac{(g+a_{{\rm c},y})L_1^2m^2}{2p_z^2}\right|\right)\nonumber\\
    &\times\mu_x\!\left(p_x-\frac{m^2a_{{\rm c},x}L_1}{p_z}\right)\mu_y\!\left(p_y-\frac{m^2(g+a_{{\rm c},y})L_1}{p_z}\right)\mu_z(p_z).
\end{align}
Here, we employed Eq.~\eqref{eq:definition_free_prop_Wigner} for the $x$- and $y$-motion and inserted the time $t=mL_1/p_z$ as well as the acceleration due to gravity and the Coriolis force.

At the first $x$-slit, the state transformation obeys Eq.~\eqref{eq:slit_trafo_classical},
\begin{align}\label{eq:jL1'}
    w'_{\rm L_1}(x,y,\blc{p})= &\frac{1}{d_{\rm s}^2}\Theta\!\left(\frac{d_{\rm s}}{2}-\left|x-x_{0{\rm s}}-L_1\frac{p_x}{p_z}+\frac{a_{{\rm c},x}L_1^2m^2}{2p_z^2}\right|\right)\Theta\!\left(\frac{d_{\rm s}}{2}-\left|y-y_{0{\rm s}}-L_1\frac{p_y}{p_z}+\frac{(g+a_{{\rm c},y})L_1^2m^2}{2p_z^2}\right|\right)\nonumber\\
    &\times\mu_x\!\left(p_x-\frac{m^2a_{{\rm c},x}L_1}{p_z}\right)\mu_y\!\left(p_y-\frac{m^2(g+a_{{\rm c},y})L_1}{p_z}\right)\mu_z(p_z)\Theta\!\left(\frac{x_{{\rm c}1}}{2}-|x-x_{01}|\right).
\end{align}
Diffraction does not play a role here since the momentum uncertainty due to the slit is typically much smaller than the width of the classical momentum distribution determined by the geometry of source and first slit. 

Since the momentum shift due to the Coriolis force $\Delta p_x=m^2a_{{\rm c},x}L_1/p_z$
is much less than the momentum spread in the source, we can approximate $\mu_x(p_x-m^2a_{{\rm c},x}L_1/p_z)\simeq\mu_x(p_x)$. Furthermore, the rightmost Heaviside function in \eqref{eq:jL1'} restricts $x$, so that the first step function restricts $p_x$. In particular, for $x_{{\rm c}1},x_{01},d_{\rm s},x_{\rm 0s}, a_{{\rm c},x}L_1^2m^2/2p_z^2\ll L_1$ the momenta are restricted to $p_x\ll p_z$. All in all, we can approximate $\mu_x(p_x-m^2a_{{\rm c},x}L_1/p_z)\simeq\mu_x(p_x)\simeq\mu_x(0)$ in Eq.~\eqref{eq:jL1'}.

After the first $x$-slit, the Wigner function propagates over a distance $L_2-L_2'$ to the first $y$-slit. There, it transforms similar to Eq.~\eqref{eq:slit_trafo_classical} before it propagates over a distance $L_2'$ to the second $x$-slit,
\begin{align}
    w_{\rm L_2}(x,y,\blc{p})= &\frac{1}{d_{\rm s}^2}\Theta\!\left(\frac{d_{\rm s}}{2}-\left|x-x_{0{\rm s}}-(L_1+L_2)\frac{p_x}{p_z}+\frac{a_{{\rm c},x}(L_1+L_2)^2m^2}{2p_z^2}\right|\right)\Theta\!\left(\frac{d_{\rm s}}{2}-\left|y-y_{0{\rm s}}-(L_1+L_2)\frac{p_y}{p_z}\right.\right.\nonumber\\
    &\left.\left.+\frac{(g+a_{{\rm c},y})(L_1+L_2)^2m^2}{2p_z^2}\right|\right)\mu_x(0)\mu_y\!\left(p_y-\frac{m^2(g+a_{{\rm c},y})(L_1+L_2)}{p_z}\right)\mu_z(p_z)\nonumber\\
    &\times\Theta\!\left(\frac{x_{{\rm c}1}}{2}-\left|x-x_{01}-L_2\frac{p_x}{p_z}+\frac{a_{{\rm c},x}L_2^2m^2}{2p_z}\right|\right)\Theta\!\left(\frac{y_{{\rm c}1}}{2}-\left|y-y_{01}-L_2'\frac{p_y}{p_z}+\frac{(a_{{\rm c},y}+g)L_2'^2m^2}{2p_z}\right|\right).
\end{align}
Here, we used that the shift of $p_x$ due to the Coriolis force is still negligible.

Diffraction at the second $x$-slit has to be taken into accounted if the first and second $x$-slit restrict the momentum spread in the beam to a value that is comparable to the momentum uncertainty due to the second slit. The state after the slit follows from Eq.~\eqref{eq:slit_trafo_quantum},
\begin{align}\label{eq:jL2'}
    w'_{\rm L_2}(x,y,\blc{p})=&\frac{1}{d_{\rm s}^2}\Theta\!\left(\frac{d_{\rm s}}{2}-\left|y-y_{0{\rm s}}-(L_1+L_2)\frac{p_y}{p_z}+\frac{(g+a_{{\rm c},y})(L_1+L_2)^2m^2}{2p_z^2}\right|\right)\mu_x(0)\mu_y\!\left(p_y-\frac{m^2(g+a_{{\rm c},y})(L_1+L_2)}{p_z}\right)\nonumber\\
    &\times\mu_z(p_z)\Theta\!\left(\frac{y_{{\rm c}1}}{2}-\left|y-y_{01}-L_2'\frac{p_y}{p_z}+\frac{(a_{{\rm c},y}+g)L_2'^2m^2}{2p_z}\right|\right)\Theta\!\left(\frac{x_{{\rm c}2}}{2}-|x-x_{02}|\right)\frac{x_{{\rm c}2}-2|x-x_{02}|}{\pi\hbar}
    \nonumber\\
    &\times\int_{-\infty}^\infty dp'\,{\rm sinc}\!\left[\frac{x_{{\rm c}2}-2|x-x_{02}|}{\hbar}(p_x-p')\right]\Theta\!\left(\frac{x_{{\rm c}1}}{2}-\left|x-x_{01}-L_2\frac{p'}{p_z}+\frac{a_{{\rm c},x}L_2^2m^2}{2p_z}\right|\right)\nonumber\\
    &\times\Theta\!\left(\frac{d_{\rm s}}{2}-\left|x-x_{0{\rm s}}-(L_1+L_2)\frac{p'}{p_z}+\frac{a_{{\rm c},x}(L_1+L_2)^2m^2}{2p_z^2}\right|\right).
\end{align}

The pinhole of the source is typically much greater than the collimating first $x$-slit, so that the source width cannot be seen from the perspective of the second $x$-slit and the illumination of the first $x$-slit can be considered as perfectly incoherent. This corresponds to neglecting the Heaviside function in the last line of Eq.~\eqref{eq:jL2'}. The associated  quantitative condition reads as
\begin{align}\label{eq:condition_infinite_source}
    d_{\rm s}>\frac{L_1}{L_2}x_{{\rm c}2}+\frac{L_1+L_2}{L_2}x_{{\rm c}1}+\left|2\frac{L_1}{L_2}x_{02}+2x_{\rm 0s}-2\frac{L_1+L_2}{L_2}x_{01}-L_1(L_1+L_2)\frac{a_{{\rm c},x}m^2}{p_z^2}\right|\,,
\end{align}
which we assume to be fulfilled in the following, so that the state no longer depends on the $x$-extension of the pinhole.

We next calculate the characteristic function \eqref{eq:definition_characteristic_funct} with respect to $(x,p_x)$, 
\begin{align}
    \chi'_{L_2}(s,y,q,p_y,p_z)=& \frac{1}{d_{\rm s}^2}
    \Theta\!\left(\frac{d_{\rm s}}{2}-\left|y-y_{0{\rm s}}-(L_1+L_2)\frac{p_y}{p_z}+\frac{(g+a_{{\rm c},y})(L_1+L_2)^2m^2}{2p_z^2}\right|\right)
    \mu_y\!\left(p_y-\frac{m^2(g+a_{{\rm c},y})(L_1+L_2)}{p_z}\right)\nonumber\\
    &\times\mu_x(0)\mu_z(p_z)\Theta\!\left(\frac{y_{{\rm c}1}}{2}-\left|y-y_{01}-L_2'\frac{p_y}{p_z}+\frac{(a_{{\rm c},y}+g)L_2'^2m^2}{2p_z}\right|\right)\chi_0(s,q), 
\end{align}
with $\chi_0(s,q)$ is given by Eq.~\eqref{eq:definitionChi}. 
It is propagated over a distance $L_3$ from the second $x$-slit to the grating,
\begin{align}
    \chi_{L_3}(s,y,q,p_y,p_z)=& \frac{1}{d_{\rm s}^2}
    \Theta\!\left(\frac{d_{\rm s}}{2}-\left|y-y_{0{\rm s}}-(L_1+L_2+L_3)\frac{p_y}{p_z}+\frac{(g+a_{{\rm c},y})(L_1+L_2+L_3)^2m^2}{2p_z^2}\right|\right)
    \mu_x(0)\mu_z(p_z)\nonumber\\
    &\times\mu_y\!\left(p_y-\frac{m^2(g+a_{{\rm c},y})(L_1+L_2+L_3)}{p_z}\right)\Theta\!\left(\frac{y_{{\rm c}1}}{2}-\left|y-y_{01}-(L_2'+L_3)\frac{p_y}{p_z}\right.\right.\nonumber\\
&\left.\left.+\frac{(a_{{\rm c},y}+g)(L_2'+L_3)^2m^2}{2p_z}\right|\right){\rm exp}\!\left[\frac{i}{\hbar}\left(\frac{a_{{\rm c},x}L_3^2m^2q}{2p_z^2}-\frac{a_{{\rm c},x}L_3m^2s}{p_z}\right)\right]\chi_0\!\left(s-\frac{L_3q}{p_z},q\right).
\end{align}
Here we employed Eq.~\eqref{eq:propagation_char_funct} for propagating in $s$ and $q$ and Eq.~\eqref{eq:definition_free_prop_Wigner} for propagating in $y$.
The state transformation at the grating the can now be carried out according to Eq.~\eqref{eq:grating_trafo_characteristic_funct},
\begin{align}
    \chi'_{L_3}(s,y,q,p_y,p_z)=&\frac{1}{d_{\rm s}^2}
   \Theta\!\left(\frac{d_{\rm s}}{2}-\left|y-y_{0{\rm s}}-(L_1+L_2+L_3)\frac{p_y}{p_z}+\frac{(g+a_{{\rm c},y})(L_1+L_2+L_3)^2m^2}{2p_z^2}\right|\right)\mu_x(0)\mu_z(p_z)\nonumber\\
    &\times\mu_y\!\left(p_y-\frac{m^2(g+a_{{\rm c},y})(L_1+L_2+L_3)}{p_z}\right)\Theta\!\left(\frac{y_{{\rm c}1}}{2}-\left|y-y_{01}-(L_2'+L_3)\frac{p_y}{p_z}\right.\right.\nonumber\\
    &\left.\left.+\frac{(a_{{\rm c},y}+g)(L_2'+L_3)^2m^2}{2p_z}\right|\right)\sum_{n=-\infty}^\infty B_n\!\left(\frac{s}{d},y\right) {\rm exp}\!\left\{\frac{i}{\hbar}\left[\frac{a_{{\rm c},x}L_3^2m^2}{2p_z^2}\left(q+\frac{2\pi\hbar}{d}n\right)-\frac{a_{{\rm c},x}L_3m^2s}{p_z}\right]\right\} \nonumber\\
    &\times \chi_0\!\left[s-\frac{L_3}{p_z}\left(q+\frac{2\pi\hbar}{d}n\right),q+\frac{2\pi\hbar}{d}n\right].
\end{align}
The explicit form of the Talbot coefficients $B_n$ for a realistic molecule-laser interaction is derived in App.~\ref{app:derivation_Talbot_coeff}. Note that the $y$-dependence of the Gaussian envelope now enters in the coefficients $B_n$ (which however do not describe diffraction in $y$-direction).

Next, we propagate the particle state over the distance $L_4-L_4'$ from the grating to the second $y$-slit, where a classical slit transformation for the $y$-motion is performed. After a final propagation 
to the screen, we find
\begin{align}
    \chi_{L_4}(s,y,q,p_y,p_z)=&\frac{1}{d_{\rm s}^2}
    \Theta\!\left(\frac{d_{\rm s}}{2}-\left|y-y_{0{\rm s}}-L\frac{p_y}{p_z}+\frac{(g+a_{{\rm c},y})L^2m^2}{2p_z^2}\right|\right)\mu_x(0)\mu_y\!\left(p_y-\frac{m^2(g+a_{{\rm c},y})L}{p_z}\right)\mu_z(p_z)
    \nonumber\\
    &\times\Theta\!\left(\frac{y_{{\rm c}1}}{2}-\left|y-y_{01}-(L_2'+L_3+L_4)\frac{p_y}{p_z}+\frac{(a_{{\rm c},y}+g)(L_2'+L_3+L_4)^2m^2}{2p_z}\right|\right)\Theta\!\left(\frac{y_{{\rm c}2}}{2}-\left|y\vphantom{\frac{(g+a_{{\rm c},y})L_4'^2m^2}{2p_z^2}}\right.\right.\nonumber\\
    &\left.\left.-y_{02}-L_4'\frac{p_y}{p_z}+\frac{(g+a_{{\rm c},y})L_4'^2m^2}{2p_z^2}\right|\right)\sum_{n=-\infty}^\infty B_n\!\left[\frac{1}{d}\left(s-\frac{L_4q}{p_z}\right),y-L_4\frac{p_y}{p_z}+\frac{(g+a_{{\rm c},y})L^2_4m^2}{2p_z^2}\right]\nonumber\\
    &\times {\rm exp}\!\left\{\frac{i}{\hbar}\left[\frac{a_{{\rm c},x}(L_3+L_4)^2m^2q}{2p_z^2}-\frac{a_{{\rm c},x}(L_3+L_4)m^2s}{p_z}+\frac{a_{{\rm c},x}L_3^2m^2}{2p_z^2}\frac{2\pi\hbar}{d}n\right]\right\}\nonumber\\
    &\times\chi_0\!\left[s-\frac{L_3+L_4}{p_z}q-\frac{L_3}{p_z}\frac{2\pi\hbar}{d}n,q+\frac{2\pi\hbar}{d}n\right].
\end{align}
We transform back to the Wigner function, and integrate over $p_x$ by using that
\begin{align}
    \int dp_x\,w_{L_4}(x,y,\blc{p})=\frac{1}{2\pi\hbar}\int_{-\infty}^\infty dq e^{-iqx/\hbar}\chi_{L_4}&(0,y,q,p_y,p_z).
\end{align}
Multiplication by $p_z$ and integration over $p_y$ and $p_z$ yields the  particle current density at the screen,
\begin{align}\label{eq:current_density_jL4_app}
    j(&x,y)\propto
    \int_{-\infty}^\infty dq\, e^{-iqx/\hbar}\sum_{n=-\infty}^\infty\int_0^\infty dp_z\, p_z\mu_z(p_z)\chi_0\!\left[-\frac{L_3+L_4}{p_z}q-\frac{L_3}{p_z}\frac{2\pi\hbar}{d}n,q+\frac{2\pi\hbar}{d}n\right]\nonumber\\
    & \times{\rm exp}\!\left\{\frac{i}{\hbar}\left[\frac{a_{{\rm c},x}(L_3+L_4)^2m^2q}{2p_z^2}+\frac{a_{{\rm c},x}L_3^2m^2}{2p_z^2}\frac{2\pi\hbar}{d}n\right]\right\}\int_{-\infty}^\infty dp_y \,h(y,p_y,p_z)B_n\!\left[-\frac{1}{d}\frac{L_4q}{p_z},y-L_4\frac{p_y}{p_z}+\frac{(g+a_{{\rm c},y})L^2_4m^2}{2p_z^2}\right].
\end{align}
The function $h(y,p_y,p_z)$ is given by Eq.~\eqref{funktionFTwoSlits}. 
The current density \eqref{eq:currentDensity} from the main text is obtained by substituting  $k=-q/\hbar$ and $\tilde{y}=y-L_4p_y/p_z+(g+a_{{\rm c},y})L_4^2m^2/2p_z^2$.

\end{widetext}

\section{Simulation parameters}\label{app:simulation_parameters_fig_1}
The parameters used to simulate the diffraction patterns in Fig.~\ref{fig:setup}(b) and Fig.~\ref{fig:fit_statistics} are motivated by a real-world experiment with phthalocyanine molecules ${\rm PcH}_2$ \cite{Simonovic2024}. The interferometer geometry is defined by the slit positions $x_{01}=-10.5{\rm \mu m}$, $x_{02}=-10.5{\rm \mu m}$, $y_{01}=0$, and $y_{02}=-15.1\,{\rm \mu m}$, slit widths $x_{\rm c1}=2.7\,{\rm \mu m}$, $x_{\rm c2}=0.6\,{\rm \mu m}$, $y_{\rm c1}=1\,{\rm m}$, and $y_{\rm c2}=20\,{\rm \mu m}$, the source position $y_{\rm 0s}=0$ and source size $d_{\rm s}=200\,{\rm \mu m}$, the grating height $y_{\rm 0g}=-3.8\,{\rm \mu m}$, and the longitudinal distances between the diffractive elements $L_1=0.52\,{\rm m}$, $L_2=0.3\,{\rm m}$, $L_2'=0.02\,{\rm m}$, $L_3=0.08\,{\rm m}$, $L_4=0.69\,{\rm m}$, and $L_4'=0.605\,{\rm m}$. The laser grating is defined by the laser power $P_{\rm L}=1\,{\rm W}$, envelope width $w_y=16\,{\rm \mu m}$, grating constant $d=133\,{\rm nm}$, laser wave length $\lambda_{\rm L}=266\,{\rm nm}$, and reflection coefficient $\eta=0.98$. The molecule is characterized by its mass $m=514.5\,{\rm u}$, polarizability $\alpha_{\rm r}=\alpha_{\rm rT}=\alpha_{\rm r0}=9\times4\pi\varepsilon_0{\rm \AA}^3$, absorption cross section $\sigma=\sigma_{\rm T}=\sigma_0=1.06{\rm \AA}^2$, quantum yields $\phi_{\rm F}=0$, $\phi_{\rm ISC}=1$, and $\phi_{\rm IC}=0$, depletion probability $P_{\rm D}=0$. The source is described by the temperature $T=746\,{\rm K}$ and the velocity shift $p_{0,z}/m=60\,{\rm m/s}$. Gravity and the Coriolis force are determined by $g=-9.81\,{\rm m/s^2}$, $\blc{\omega}\cdot\blc{e}_x=5.4\times10^{-5}{\rm /s}$, and $\blc{\omega}\cdot\blc{e}_y=-4.9\times10^{-5}{\rm /s}$. The screen is discretized into squares of edge length $d_{\rm px}=0.33\,{\rm \mu m}$. It consists of 1004 pixels in $x$-direction and 1003 in $y$-direction. 

In  Fig.~\ref{fig:diffraction_patterns}, the interferometer geometry is
defined by $x_{01}=0$, $x_{02}=0$, $x_{\rm c1}=3\,{\rm \mu m}$, $x_{\rm c2}=2\,{\rm \mu m}$, $y_{\rm 0s}=0$, $d_{\rm s}=100\,{\rm \mu m}$, $y_{01}=0$, $y_{02}=-15.7\,{\rm \mu m}$, $y_{\rm c1}=1\,{\rm m}$, $y_{\rm c2}=10\,{\rm \mu m}$, $y_{\rm 0g}=-4.3\,{\rm \mu m}$, $L_1=0.5\,{\rm m}$, $L_2=0.3\,{\rm m}$, $L_2'=0.02\,{\rm m}$, $L_3=0.2\,{\rm m}$, $L_4=0.7\,{\rm m}$, and $L_4'=0.6\,{\rm m}$. 
The laser grating parameters 
are as above ($P_0=1$\,W) except for $\eta=1$. 
Also the source and  molecular parameters are as above, except for
$\alpha_{\rm r,T}=\alpha_{\rm r}=\alpha_0=25\times4\pi\varepsilon_0{\rm \AA}^3$, $\sigma_{\rm T}=\sigma=\sigma_0=0.7{\rm \AA}^2$.

\bibliography{bibliography}
 \newpage

\end{document}